\documentclass[aps,prd,twocolumn,nofootinbib,eqsecnum,superscriptaddress,longbibliography]{revtex4-2}

\usepackage[centertags]{amsmath}
\usepackage{amssymb}
\usepackage{latexsym}
\usepackage{enumerate}
\usepackage{graphicx}
\usepackage{mathrsfs}
\usepackage[hidelinks]{hyperref}
\usepackage{stmaryrd}
\usepackage{import}
\usepackage{tensor}
\usepackage[normalem]{ulem}
\usepackage{xcolor}
\usepackage{bm}
\usepackage{multirow}
\usepackage{breakurl}
\usepackage{float}
\usepackage{lipsum}
\usepackage{siunitx}
\usepackage{comment}
\usepackage{mathtools}
\usepackage{autobreak}
\usepackage{subfig}
\usepackage{orcidlink}
\mathtoolsset{centercolon}
\usepackage{comment}

\definecolor{grey}{rgb}{0.4,0.4,0.4}
\definecolor{dullmagenta}{rgb}{0.4,0,0.4}
\definecolor{darkblue}{rgb}{0,0,0.4}
\definecolor{midblue}{rgb}{0,0,0.5}
\definecolor{midred}{rgb}{0.5,0,0}
\definecolor{orange}{rgb}{1,0.5,0}
\definecolor{lightbrown}{rgb}{0.75,0.5,0.25}
\definecolor{tan}{cmyk}{0.14,0.42,0.56,0}
\definecolor{djunglegreen}{cmyk}{0.99,0,0.52,0}
\definecolor{lightgreen}{rgb}{0,1,0}
\definecolor{olivegreen}{cmyk}{0.64,0,0.95,0.40}
\definecolor{midgreen}{rgb}{0.0,0.675,0.0}
\definecolor{darkgreen}{rgb}{0,0.5,0}
\definecolor{ceruleanblue}{rgb}{0.0, 0.45, 1} 
\definecolor{burgundy}{rgb}{0.5, 0.0, 0.13}
\definecolor{hvred}{RGB}{186,12,47}
\definecolor{ste}{rgb}{0.01, 0.28, 1.0}

\hypersetup{
    colorlinks=true,
    linkcolor=ceruleanblue,
    filecolor=ceruleanblue,      
    urlcolor=ceruleanblue, 
    citecolor=midblue, 
}

\include{preamble}

\allowdisplaybreaks

\begin{document}

\title{Analytical calculation of self-force effects on a scalar particle in an eccentric orbit around a Schwarzschild black hole}

\begin{abstract}
   In this work, we analytically investigate the effects of the scalar self-force exerted by a massless scalar field on a particle in a slightly eccentric orbit around a Schwarzschild black hole. By solving the Klein-Gordon equation in the curved spacetime background, using a combination of post-Newtonian (PN) expansion, and small-eccentricity approximation, we derive explicit expressions for the self-force components at the particle location, as well as for the associated energy and angular momentum fluxes. Our results are valid up to sixth post-Newtonian (6PN) order and fourth order in eccentricity ($e^4$). We  compare asymptotic fluxes with those obtained in \href{https://doi.org/10.1103/PhysRevD.109.104003}{Phys. Rev. D 109, 104003 (2024)} for scalar-tensor (ST) theories. Once the relation between the two approaches has been established, we find perfect agreement by fixing the asymptotic value of the scalar field in ST theory $\phi_0 = 1$. 
\end{abstract}

\author{Salvatore Capozziello\orcidlink{0000-0003-4886-2024}}\email{capozziello@na.infn.it}\affiliation{Scuola Superiore Meridionale, Largo San Marcellino 10, 80138, Naples, Italy}\affiliation{INFN, Sezione di Napoli, Complesso Universitario di Monte S. Angelo, Via Cinthia Edificio 6, 80126, Naples, Italy}\affiliation{Dipartimento di Fisica ”E. Pancini”, Università di Napoli “Federico II”, Via Cinthia Edificio 6, I-80126, Napoli, Italy}

\author{Nicola Menadeo\orcidlink{0009-0002-8577-4592}}\email{nicola.menadeo@aei.mpg.de}\affiliation{Scuola Superiore Meridionale, Largo San Marcellino 10, 80138, Naples, Italy}\affiliation{INFN, Sezione di Napoli, Complesso Universitario di Monte S. Angelo, Via Cinthia Edificio 6, 80126, Naples, Italy}
\affiliation{Max Planck Institute for Gravitational Physics (Albert Einstein Institute) \\
Am Mühlenberg 1, D-14476 Potsdam-Golm, Germany}

\author{Davide Usseglio\orcidlink{0000-0003-2427-9547}}
\email{davide.usseglio-ssm@unina.it}
\affiliation{Scuola Superiore Meridionale, Largo San Marcellino 10, 80138, Naples, Italy}
\affiliation{INFN, Sezione di Napoli, Complesso Universitario di Monte S. Angelo, Via Cinthia Edificio 6, 80126, Naples, Italy}

\maketitle

\section{Introduction}
The groundbreaking detection of gravitational waves (GWs) by the LIGO and Virgo collaborations \cite{LIGOScientific:2016aoc, LIGOScientific:2017vwq}  opened a new avenue for investigating gravity in extreme regimes, particularly around black holes. This discovery has not only confirmed the theoretical foundations of General Relativity (GR), but also led to constraining a wide range of alternative gravity theories, many of which have been systematically ruled out~\cite{Ezquiaga:2017ekz, Creminelli:2017sry}. The next generation of GW observatories, including upgrades to current detectors and new facilities such as LISA, the Einstein Telescope and Cosmic Explorer~\cite{LISA:2017pwj, Punturo:2010zz,Gupta:2023lga,Evans:2021gyd}, will significantly enhance the ability to detect potential deviations from GR~\cite{Berti:2015itd,Ezquiaga:2018btd,Yunes:2013dva,Lagos:2024boe,LIGOScientific:2018dkp,Bellini:2017avd}.

One way to extend Einstein's theory is to introduce an additional scalar field alongside the metric tensor. Depending on the nature of the coupling between the scalar field and the gravitational field, the scalar degree of freedom acts as a source term that does not change the underlying gravitational theory or it could lead to genuine modifications of the gravitational interaction, while preserving many of the fundamental symmetries of the original theory; this class of gravity models is known as Scalar–Tensor (ST) theories, in which gravity is mediated jointly by the usual spin-2 graviton and a spin-0 scalar mode. Over the years, such models have offered a flexible framework for probing deviations from standard GR, with applications mainly in the cosmological setting, i.e. for early (and late) time cosmic acceleration~\cite{Cusin:2017mzw,Khoury:2003rn,Nakamura:2020ihr,Salzano:2017qac,Lombriser:2015sxa,Sakstein:2017xjx,Li:2012dt}. 

However, ST can also be employed in modeling astrophysical systems. In particular, there has been a revival in the investigation of ST theories in the generation of GWs from binary systems: among them, the Brans-Dicke (BD) model~\cite{jordan1955schwerkraft, Fierz:1956zz} offers the simplest extension of GR by introducing a dynamical scalar field. In such a scenario, substantial efforts have been devoted to parametrize the impact of additional fields on observables, both analytically~\cite{Bernard:2018hta,Bernard:2018ivi,Bernard:2022noq,Bernard:2023eul,Trestini:2024zpi,Damour:1996ke,Trestini:2024mfs,Capozziello:2005bu,Jain:2022nxs,Jain:2023fvt} and numerically~\cite{Shibata:2013pra,Julie:2022qux}, with the aim of assessing whether such effects could be detectable in upcoming GW observations. These features could not only be due to a modification of the underlying gravitational theory, but they might also be interpreted as environmental effects on the binary, in which the scalar field plays the role of surrounding matter that interacts with the binary (see~\cite{Dyson:2024qrq,Dyson:2025dlj}).

For the purpose of assessing the effects of scalar fields, intended either as a true modification to Einstein's gravity or as an environmental effect, on the GW signal, Extreme Mass Ratio Inspiral (EMRI) events have emerged as powerful probes of fundamental physics~\cite{Barsanti:2022ana,Spiers:2023cva,DellaRocca:2024pnm}, especially when analyzed within the sensitivity range of LISA~\cite{Babak:2017tow,Speri:2024qak,Singh:2026sad,Strub:2025dfs}. EMRIs can be naturally investigated within the Self Force (SF) formalism \cite{Poisson:2011nh,Pound:2021qin}, where the central black hole of mass $M$, labeled  \textit{primary}, describes the background geometry, while the \textit{secondary} object can be treated as a localized small perturbation of the background metric with mass $\mu$, where $\mu/M\ll1$.

In this work, we employ tools from Black Hole Perturbation Theory (BHPT)~\cite{Poisson:2011nh,Pound:2021qin}, which provides a simplified yet insightful framework to study how the evolution of a scalar field is affected by the background geometry. Specifically, we focus on the Scalar Self-Force (SSF), which arises when a scalar charge interacts with its own field while moving along a prescribed trajectory. In this setup, we aim to analytically investigate the dynamics of a scalar particle in an equatorial, eccentric orbit around a Schwarzschild black hole, which has already been studied in~\cite{Vega:2013wxa} only via numerical techniques. Our calculations will be valid only at first order in $\mu/M\ll1$, where $\mu$ is the mass of the scalar particle and $M$ is the mass of the Schwarzschild black hole.

The ultimate goal of this analysis is to evaluate the components of the self-force. We then compute the scalar radiation via energy fluxes, which will provide a natural consistency check for our self-force calculation. Therefore, we neglect the backreaction of the scalar field on the background metric, as it would represent a higher-order effect in the perturbative expansion.  
In order to achieve this, we have to solve the differential equation governing the scalar field evolution, i.e. the spin-$s=0$ Teukolsky equation, which, in the non-spinning case, reduces to the Klein–Gordon equation on a Schwarzschild background, sourced by a point-like particle. Similar setups have been widely explored in the gravitational self-force (GSF) context for $|s|=2$ perturbations~\cite{Bini:2015bfb,Bini:2016qtx,Kavanagh:2015lva}, including extensions to rotating black holes~\cite{Bini:2016dvs,Kavanagh:2017wot,Bini:2019lcd}, and have provided valuable input for improving the Effective One Body (EOB) framework~\cite{Bini:2016cje,Bini:2018aps}. Analytical calculations in this setting have also reached very high orders in small-eccentricity expansions~\cite{Munna:2020iju,Munna:2022gio,Munna:2022xts,Munna:2023wce}.

The paper is organized as follows.
In Sec.~\ref{sec:2} we present a review of the eccentric orbits in Schwarzschild in the small eccentricity limit. We show explicit expressions for both the geodesic motion, periastron advance and radial orbital period. After these preliminaries, we provide a detailed discussion in Sec.~\ref{sec_3} on how to solve the Klein-Gordon equation in the Schwarzschild background with a source that is forced to be on an eccentric orbit. We also provide a comparison of our results against the circular limit, by looking at the impact of the eccentricity at the various orders in the PN expansion. In Sec.~\ref{sec_4} we compute the SF components acting on the scalar particle, showing the differences compared to the circular limit. In Sec.~\ref{sec_5} we complete our analytical computations with the evaluation of the asymptotic fluxes of energy and angular momentum, subsequently compared, in Sec.~\ref{sec_6}, with the scalar fluxes in Ref.~\cite{Trestini:2024zpi} by establishing the correct mapping between the two sets of results.

As a side note, the Authors were surprised to note that analytical expressions for the scalar eccentric case were never investigated in detail in the literature, even in light of the recent interest mentioned in the previous paragraph. Another reason for this work was to close this gap.

\textit{Conventions.} We use $G=c=1$. We also use a placeholder $\eta=1/c$ to clarify for the reader the nature of some PN series. Further, we use $\log$ to indicate the natural logarithm. The signature for the Schwarzschild metric is $(-,+,+,+)$.

\section{Eccentric geodesic motion} \label{sec:2}

Let us consider a Schwarzschild background geometry with the following line element 
\begin{equation}
    ds^2 = -f(r)dt^2 + \dfrac{dr^2}{f(r)} + r^2 (d\theta^2 + \sin^2\theta d\phi),
\end{equation}
in the usual set of Schwarzschild coordinates $(t,r,\theta,\phi)$ and $f(r) = 1-2M/r$. Generic timelike trajectories $x^\mu=x^\mu_p(\tau)$ in the equatorial plane $(\theta=\pi/2)$ can be parametrized in terms of the proper time $\tau$. 

According to these definitions, the components of the four-velocity $u^\mu = \dot{x}^\mu$ can be expressed in terms of the (specific) conserved energy $E$ and angular momentum $L$ as follows:
\begin{align}
    \dot{t}_p = \dfrac{E}{f(r_p)},\hspace{0.2cm}
    \dot{r}^2_p = E^2 - V(r_p,L),\hspace{0.2cm}
    \dot{\phi}_p = \dfrac{L}{r^2_p},\label{geod_eq}
\end{align}
where 
\begin{equation}
    V(r_p,L) = f(r_p)\left(1+\dfrac{L^2}{r^2_p}\right),
\end{equation}
and the dot indicates the derivative with respect to $\tau$.

By solving the system in Eq.~\eqref{geod_eq}, we can write the components of the geodesic in terms of energy and angular momentum. However, since we are interested in eccentric orbits, it is more convenient to change the parameterization and to use the semi-latus rectum $p$ and the (Darwin) eccentricity $e$ by employing the expressions
\begin{equation}\label{def:energy_angmom}
    E^2 = \dfrac{(p-2)^2 - 4e^2}{p(p-3-e^2)} \,, L^2 = \dfrac{p^2 M^2}{p-3-e^2},
\end{equation}
where we recall that $p>6+2e$ and $0\leq e<1$ for bound orbits.

With this set of variables, we can write the radial motion straightforwardly as
\begin{equation}\label{eq:geod:radial}
    r_p(\chi) = \dfrac{M p}{1+e\cos\chi},
\end{equation}
with $\chi$ as the usual relativistic anomaly that takes values between $0$ and $2\pi$, which corresponds to the periastron and apoastron of the eccentric orbit. The geodesic equations can then be written in terms of the parameter $\chi$ as
\begin{subequations}\label{eq:geod_eq_chi}
    \begin{align}
    \dfrac{d\phi}{d\chi} &= \dfrac{\sqrt{p}}{\sqrt{p-6-2e\cos\chi}}\,,\\
    \dfrac{dt}{d\chi} & =  \frac{M p^2}{(p-2-2e\cos\chi)(1+e\cos\chi)^2} \sqrt{\frac{(p-2)^2-4e^2}{p-6-2e\cos\chi}}\,,
\end{align}
\end{subequations}
where the solutions for both equations can be written in terms of elliptic integrals, i.e.
\begin{equation}\label{angular_var_full}
    \phi_p(\chi) = \left(\dfrac{4p}{p-6-2e}\right)^{1/2}F\left( \dfrac{\chi}{2} \bigg| -\dfrac{4e}{p-6-2e} \right)\,,
\end{equation}
for the angular variable, with initial condition $\phi_p(0)=0$. It is also possible to get a solution of this form for $t_p(\chi)$ but the procedure is more involved and it is not relevant for our discussion, though the expression can be found in \cite{Fujita:2009bp}. Other representations for the solutions of geodesic motion in the Schwarzschild spacetime can be found, see for instance \cite{Scharf:2011ii,Cieslik:2022uki}. However, it is more useful to produce the geodesic periastron advance and the (radial) orbital period as
\begin{equation}
    \Delta\phi = 2\int_0^\pi d\chi\,\dfrac{d\phi}{d\chi} = \frac{4 \sqrt{p}}{\sqrt{-6-2 e+p}}\mathcal{K}\left(\frac{4 e}{6+2 e-p}\right)
\end{equation}
\begin{align}\nonumber
    \Delta T_r  = 2\int_0^\pi d\chi\,\dfrac{dt}{d\chi} & = C_1(p,e) \mathcal{K}\left(\frac{4 e}{-6+2 e+p}\right) \\ \nonumber
    & \hspace{-1cm} +C_2(p,e) \mathcal{E}\left(\frac{4 e}{-6+2
   e+p}\right)\\ \nonumber
    & \hspace{-1cm}+C_3(p,e) \Pi \left(\frac{2 e}{-1+e},\frac{4 e}{-6+2
   e+p}\right) \\
   & \hspace{-1cm} +C_4(p,e) \Pi \left(\frac{4 e}{-2+2 e+p},\frac{4 e}{-6+2
   e+p}\right) ,
\end{align}
where $\mathcal{E}(\cdot)$, $\mathcal{K}(\cdot)$ and $\Pi(\cdot)$ are the complete elliptic integrals of the first, second and third kind, respectively. Further, the $C_i$ are rational functions of $(p,e)$.

Eccentric orbits have two fundamental frequencies which are associated with the libration between periastron and apoastron ($\Omega_r$) and the average rate of azimuthal advance over one radial period ($\Omega_\phi$), which are respectively defined as
\begin{equation}
    \Omega_r \equiv \dfrac{2\pi}{\Delta T_r},\,\,\, \Omega_\phi \equiv \dfrac{\Delta\phi}{\Delta T_r}\,.
\end{equation}

Let us discuss how to obtain perturbative expressions for the aforementioned quantities. In particular, we are interested in the post-Newtonian (PN) approximation, which can be performed using $u_p=1/p$ as PN expansion parameter. By implementing this expansion in the fundamental frequencies it is possible to check that $\Omega_r =\Omega_\phi$ at the Newtonian level. However, it is more convenient to use another counting parameter for the PN expansion, namely
\begin{equation}
    y = (M\Omega_\phi)^{2/3}\,,
\end{equation}
from which one can obtain the expression of $p$ in terms of $y$ by inverting the series in $u_p$ of $\Omega_\phi$. 

Once the PN expansion has been implemented on the various geodesic quantities, we also want to take into account a small eccentricity expansion. By expanding Eqs.~\eqref{eq:geod:radial} and~\eqref{angular_var_full}, one finds
\begin{align}\label{eq:rad_geod} \nonumber
    \dfrac{r_p(\chi)}{M} & = \frac{1}{y}-\frac{e \cos (\chi )}{y} -\frac{e^2}{y} \left[\sin ^2(\chi ) + P_2(y) \right] \\ \nonumber 
    &  +\frac{e^3}{ y} \cos (\chi ) \left[ \sin^2 ( \chi )+P_3(y)\right] \\ 
    & +\frac{e^4}{8 y}\left[ \cos(4\chi) + P_{4}(y) + Q_4(y)\cos(2\chi)\right] + O(e^5) \,,
\end{align}
\begin{align} \nonumber
   \phi_p(\chi) & =  R_0(y)\chi + e \, R_1(y) \sin(\chi) \\ \nonumber
   & + e^2 \left[ R_2(y) \chi + S_2(y) \sin(2\chi) \right] \\ \nonumber
   & + e^3 \left[ R_3(y) \sin(\chi) + S_3(y) \sin(3\chi)  \right] \\ 
   & + e^4 \left[ R_4(y)\chi + S_4(y) \sin(2\chi)+ T_4(y) \sin(4\chi) \right] \,,
\end{align}\label{eq:ang_geod}
up to $e^4$ and where the $P_i(y)$, $Q_i(y)$, $R_i(y)$, $S_i(y)$ and $T_i(y)$ are polynomials in $y$ that could be computed at arbitrary PN order. While the radial motion is explicitly periodic, the expression of $\phi_p(\chi)$ has also some linear-in-$\chi$ contributions which take into account that this quantity should not be periodic. It is useful to look at such a split also in the time domain expression for $\phi_p(t)$, which can be schematized as \cite{Barack:2008ms,Hopper:2015icj}
\begin{equation}
    \phi_p(t) = \Omega_\phi t + \Delta \phi(t)\,,
\end{equation}
where the first term represents the mean azimuthal advance, while the second is periodic over a radial period and, by reintroducing $\chi$ in this expression one gets 
\begin{subequations}
    \begin{align}\nonumber
    \Omega_\phi t(\chi) &=P'_0(y)\chi + e P'_1(y)\sin(\chi) \\ \nonumber
    & \hspace{-0.2cm} + e^2 \left[ P'_2(y)\chi + Q'_2(y)\sin(2\chi) \right] \\  \nonumber
    & \hspace{-0.2cm} + e^3 \left[ P'_3(y)\sin(\chi) + Q'_3(y)\sin(3\chi) \right] \\ 
    & \hspace{-0.2cm} + e^4 \left[ P'_4(y)\chi + Q'_{4,2}(y)\sin(2\chi) + Q'_{4,4}(y)\sin(4\chi) \right], \\ \nonumber
    \Delta\phi(\chi) &= R'_1(y)e\sin(\chi)+R'_2(y)e^2\sin(2\chi) \\ \nonumber 
    & + e^3 \left[ R'_3(y)\sin(\chi) + S'_3(y)\sin(3\chi) \right] \\ 
    & + e^4 \left[ R'_4(y)\sin(2\chi) + S'_4(y)\sin(4\chi) \right] \,,
\end{align}
\end{subequations}
which shows that the first line does not contain only linear-in-$\chi$ contributions but also some trigonometric functions, while the second line presents only $\sin(n\chi)$ with $|n|=1,2,3,4$. Again, we have introduced some polynomials in $y$ to ease the notation and highlight the structure in $\chi$. This separation in the time domain will turn out to be important in the next Section, where we will solve the Klein-Gordon equation sourced by a scalar charge on such orbits.

\section{Scalar Field in eccentric motion around a Schwarzschild black hole}\label{sec_3}

Let us now assume that a massless scalar field $\psi$, sourced by a scalar charge with mass $\mu\ll M$, is orbiting around the Schwarzschild black hole following an eccentric geodesic. The scalar field can then be modeled as a small perturbation to the Schwarzschild background, which obeys the spin-0 Teukolsky equation \cite{Bini:2024icd}. The equation reduces to the massless Klein-Gordon equation on a Schwarzschild background
\begin{equation}\label{KG-eq}
    \Box \psi = 4\pi \rho,
\end{equation}
where $\rho$ is the charge density with support only along the particle world line
\begin{align}\label{def:source}
    \rho(x^\mu) = q \int (-g)^{-1/2} \delta^{(4)}(x^\mu-x_p^\mu(\tau)) d\tau\,
\end{align}
and $\Box$ is a second-order differential operator defined as
\begin{equation}
    \Box = \dfrac{1}{\sqrt{-g}}\partial_\mu (\sqrt{-g}g^{\mu\nu} \partial_\nu).
\end{equation}

By leveraging the spherical symmetry of the Schwarzschild background, the field $\psi$ can be decomposed in terms of the (scalar) spherical harmonics $Y_{lm}(\theta,\phi)$ basis as follows
\begin{align}\label{eq:scal_field}
    \psi (x^\mu) = \sum_{l,m}\psi_{lm}(t,r) Y_{lm}(\theta,\phi),
\end{align}
where we have separated the radial and the angular dependence\footnote{The equation under study is completely separable and the solutions to the angular equation are indeed the $Y_{lm}(\theta,\phi)$.}. This allows us to solve the radial and the angular equations separately by keeping the indices $(l,m)$ as parameters. Because we are interested in obtaining analytical expressions for $\psi(x^\mu)$, we need to expand each $\psi_{lm}(t,r)$ in Fourier series as
\begin{equation}
    \psi_{lm}(t,r) = \sum_{n=-\infty}^{\infty} \psi_{lmn}(r)e^{-i\omega t}\,,
\end{equation}
where $\omega\equiv\omega_{mn} = m\Omega_\phi + n\Omega_r$, which is a consequence of the bi-periodic motion of the source. The Fourier coefficients can be obtained by using standard definitions as
\begin{equation}
    \psi_{lmn}(r) = \dfrac{1}{\Delta T_r}\int_{0}^{\Delta T_r} dt\, e^{i\omega t}  \psi_{lm}(t,r)\,.
\end{equation}
Indeed, the same discussion can be done for the source term $\rho(x^\mu)$ to get the coefficients of the Fourier series $\rho_{lm}(t,r)$ as
\begin{align} \nonumber
    \rho_{lmn}(r) & = \dfrac{q}{\Delta T_r\,E}Y^*_{lm}\left(\dfrac{\pi}{2},0\right)\\
    &\times \int_0^{\Delta T_r} dt \, \dfrac{f(r(t))}{r(t)^2} \delta(r'-r(t))  e^{i(\omega t-m\phi_p(t))}\,,\label{eq:mode-decomposed charge density}
\end{align}
where we derived the integrand from Eq.~\eqref{def:source}.

By using these mode-decomposed quantities, the equation for the scalar field $\psi_{lmn}(r)$ in Fourier space assumes the following form
\begin{equation} \label{eq_KG_fourier}
    \mathcal{L}_r \, \psi_{lmn}= -4\pi r^{-2}\rho_{lmn},
\end{equation}
where
\begin{align} 
   \mathcal{L}_r & \equiv  \dfrac{d^2}{dr^2} + \dfrac{2(r-M)}{r^2 f(r)}\dfrac{d}{dr}
      + \left[ \dfrac{\omega^2}{f(r)^2}-\dfrac{l(l+1)}{r^2f(r)} \right]  \,,
\end{align}
is a second-order differential operator. 

The general inhomogeneous solution can be written for each $(l,m)$-modes as 
\begin{equation}\label{eq:scalar_field_freqdomain}
\psi_{lmn}(r)=-4\pi\int dr' G_{lmn}(r,r') r'^2 \rho_{lmn}(r'),
\end{equation}
where $G_{lmn}(r,r')$ is the Green function defined in terms of the homogeneous solutions
\begin{align}\label{def:GF}\nonumber
    G_{lmn}(r,r') & = \dfrac{1}{W_{lmn}}\big[ R_{\rm{in}}^{lmn}(r)R_{\rm{up}}^{lmn}(r') H(r'-r) \\
    & + R_{\rm{in}}^{lmn}(r')R_{\rm{up}}^{lmn}(r) H(r-r') \big],
\end{align}
$H(\cdot)$ is the Heaviside step function, $R_{\rm{in}}^{lmn}(r)$ and $R_{\rm{up}}^{lmn}(r)$ are two independent homogeneous solutions of the radial equation with the correct boundary conditions at the event horizon of the Schwarzschild black hole and at infinity, respectively. Further, $W_{lmn}$ is the invariant Wronskian
\begin{equation}
    W_{lmn} = r^2 f(r) \left[ R_{\rm{in}}^{lmn}(r) R_{\rm{up}}'^{lmn}(r) - R_{\rm{up}}^{lmn}(r) R_{\rm{in}}'^{lmn}(r) \right].
\end{equation}

By plugging \textcolor{black}{Eq.~\eqref{eq:mode-decomposed charge density}} into Eq.~\eqref{eq:scalar_field_freqdomain} one has
\begin{align}\nonumber
    \psi_{lmn}(r) & = -\dfrac{4\pi q}{\Delta T_r E} Y^*_{lm}\left( \dfrac{\pi}{2},0 \right)\int dr' G_{lmn}(r,r') r'^2 \\
    & \times \int_{0}^{\Delta T_r} dt \, \dfrac{f(r(t))}{r(t)^2} \delta(r'-r(t))  e^{i(\omega t-m\phi_p(t))}\,, 
\end{align}
and the integration over $r'$ becomes trivial because of the delta function, hence one gets
\begin{align}\nonumber
    \psi_{lmn}(r) & = -\dfrac{4\pi q}{\Delta T_r E} Y^*_{lm}\left( \dfrac{\pi}{2},0 \right)  \\
    & \times  \int_{0}^{\Delta T_r} dt\, G_{lmn}(r,r_p(t)) f(r(t))   e^{i(\omega t-m\phi_p(t))}\,.
\end{align}

It is now convenient to recall the definition of $\omega$ in terms of the two fundamental frequencies of the eccentric motion and the time domain expression of $\phi_p(t)$ which leads to
\begin{align}\nonumber
    \psi_{lmn}(r) & = -\dfrac{4\pi q}{\Delta T_r E} Y^*_{lm}\left( \dfrac{\pi}{2},0 \right)  \\
    & \times  \int_{0}^{\Delta T_r} dt\, G_{lmn}(r,r_p(t)) f(r(t))   e^{i (n\Omega_r t -m\Delta \phi(t))}\,,
\end{align}
and now, by changing the integration variable to $\chi$ as
\begin{align}\nonumber
    \psi_{lmn}(r) & = -\dfrac{4\pi q}{\Delta T_r E} Y^*_{lm}\left( \dfrac{\pi}{2},0 \right)  \\
    & \hspace{-0.7cm} \times  \int_{0}^{2\pi} d\chi \, \dfrac{dt}{d\chi} G_{lmn}(r,r_p(\chi)) f(r_p(\chi))   e^{i (n\Omega_r t -m\Delta \phi(\chi))}\,,
\end{align}
it becomes clear that the integrand is periodic over an orbital cycle. One only has to perform the remaining integral, which is however very simple and involves only the integration of complex exponentials. 

Once the integral is performed, there are three summations that need to be exploited: the sum over $n$ which produces the time domain expression $\psi_{lm}(t,r)$, which is followed by the sums over $(l,m)$.

Since we are interested in a small eccentricity expansion, the infinite sum over $n$ actually truncates at a finite number. In particular, at the specific order $e^i$, we have that the only non-zero terms of this sum are the ones with $|n|\leq i$. Besides, it is convenient to write the time domain $(l,m)$-modes of the field as
\begin{align}\nonumber
    \psi_{lm}(t,r) & = e^{-im\Omega_\phi t}\sum_n \psi_{lmn}(r) e^{-in\Omega_r t} \\
    &= e^{-im\Omega_\phi t} \bar{\psi}_{lm}(t,r)\,, \label{eq:scal_field_timedom}
\end{align}
this choice will become clear in the next paragraphs.

The final steps involve the summation over $(l,m)$-modes but, more crucially, the evaluation of the time domain field $\psi(t,r,\theta,\phi)$ at the particle's position, i.e. $(r,\theta,\phi)\to(r_p(t),\pi/2,\phi_p(t))$. The resulting expression is, however divergent and must be suitably regularized through the mode-sum regularization \cite{Barack:1999wf,Barack:2009ux}. This method is well motivated, as Eq.~\eqref{eq:scal_field} corresponds to the retarded field $\psi^{\text{ret}}(t,r,\theta,\phi)$, which is divergent at the particle location by construction, while its  individual spherical-harmonic modes remain finite. The regularized field $\psi_{\text{R}}(t)$ can be thus computed by subtracting from each  $l$-mode a suitable regulator $B(t)$ which does not depend on $l$. 

The first step is to sum over $m$ as
\begin{align}\nonumber
    \psi^{\text{ret}}_l (t) &= \sum_{m=-l}^l\psi_{lm}(t,r)Y_{lm}(\theta,\phi)\Bigg|_{(r,\theta,\phi)\to(r_p(t),\pi/2,\phi_p(t))}\\
    & = \sum_{m=-l}^l  \bar{\psi}_{lm}(t,r) Y_{lm}(\pi/2,0)e^{im \Delta\phi(t)} \bigg|_{r\to r_p(t)} \,,
\end{align}
where the summation can be straightforwardly performed by using standard formulas (see Appendix F of Ref.~\cite{Nakano:2003he}). The retarded field is now explicitly periodic because the only non-periodic term produced by $e^{-im\Omega_\phi t}$ simplifies when the retarded field is evaluated at the particle position. Hence, the regular field is computed by subtracting the regulator \textit{before} the summation over $l$
\begin{equation}
    \psi_{\text{R}}(t) = \sum_l\left\{ \psi^{\text{ret}}_l (t) - B(t) \right\}.
\end{equation}

To obtain explicit expressions for the regularized field, one must first express the Green function in terms of the homogeneous solutions $R_{\rm{in}}^{lm\omega}(r)$ and $R_{\rm{up}}^{lm\omega}(r)$ appearing in Eq.~\eqref{def:GF}. The most straightforward approach consists in solving Eq.~\eqref{eq_KG_fourier} expanding in PN and working out the equation order-by-order.  This procedure yields what we refer to as the
\textit{PN solution}, which can be written for a generic $l$-mode. However, it does not satisfy the physical boundary condition at the horizon. This behaviour is reflected in the presence of some divergences at specific PN-order for certain values of $l$. To address this issue, the most commonly used technique is the so-called Mano-Suzuki-Takasugi method (MST) \cite{Sasaki:2003xr}, which relies on writing the homogeneous solutions for the full equation in terms of an infinite sum of hypergeometric functions: 
these are referred to as the \textit{MST solutions}. In practice, computing observables up to some $n_{\text{max}}$-PN, requires including all the MST solutions up to some $l\leq l_{\text{max}}$\footnote{The $l_{\text{max}}$ can be deduced by looking at the coefficient in the PN solution of the specific order we are interested in and check for which values of $l$ it diverges. We identify the largest one as $l_{\text{max}}$.}. The downside of this procedure is that each $l$-mode needs to be computed independently.

Besides, the MST solutions show an interesting analytical structure. In particular, the divergence in $l$ of the PN solutions is translated in $\log(-i\omega r)$ which is indeed a problem in the high frequency limit. This is not a real issue for bound orbits, such as circular or eccentric, where the frequency spectrum is discrete and the large-frequency limit is never encountered in the calculation. A very different situation appears in the hyperbolic scattering case where the high frequency limit plays a crucial role and a suitable regularization procedure should be taken into account to extract physical quantities from the calculation (see Ref.~\cite{Bini:2024icd}).

\begin{figure*}[t]
  \centering
  \includegraphics[width=\linewidth]{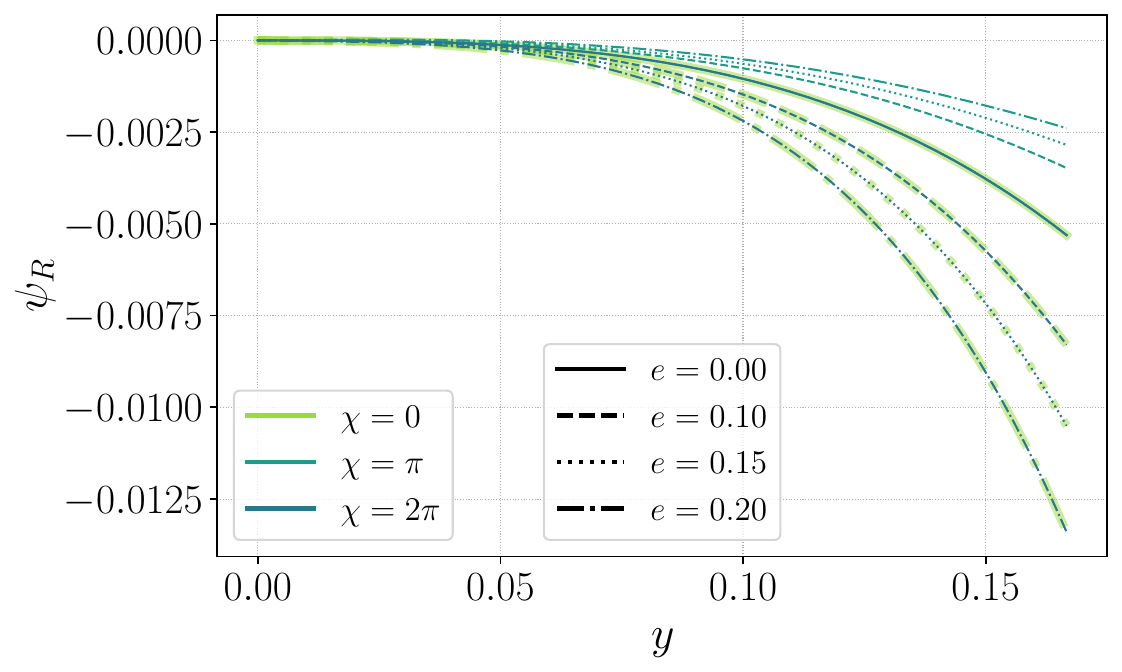}
  \caption{Regularized field $\psi_R(y, e; \chi)$ evaluated at $\chi = 0, \pi , 2\pi$, setting $M=1$.}
  \label{fig:Fig_psiR}
\end{figure*}

It is now possible to present the expressions for the parameter $B(\chi)$ and the regularized field, respectively; the periodic structure is well preserved at each order in $e$. Hence, up to 6PN and $e^4$ (in units of $q$) 
\begin{widetext}
\begin{align}\nonumber
 B(y,e;\chi) & =  y-\frac{y^2}{4}-\frac{39 y^3}{64}-\frac{385 y^4}{256}-\frac{61559
   y^5}{16384}-\frac{622545 y^6}{65536} \\ \nonumber 
   & + e \left(y-\frac{3 y^2}{4}-\frac{99 y^3}{64}-\frac{823
   y^4}{256}-\frac{110511 y^5}{16384}-\frac{938331 y^6}{65536}\right)
   \cos \chi  \\ \nonumber 
   & + e^2 \bigg[y-\frac{23 y^2}{8}-\frac{15 y^3}{2}-\frac{12677
   y^4}{512}-\frac{1758353 y^5}{16384}-\frac{73637235
   y^6}{131072} \\ \nonumber 
   & \hspace{0.5cm}-\left(\frac{3 y^2}{8} +\frac{27
   y^3}{64}+\frac{93 y^4}{512}-\frac{8001 y^5}{8192}-\frac{612873
   y^6}{131072}\right) \cos 2 \chi \bigg] \\ \nonumber
   & + e^3 \cos \chi  \bigg[y-\frac{29 y^2}{8}-\frac{129 y^3}{16}-\frac{10387
   y^4}{512}-\frac{1154065 y^5}{16384}-\frac{44917209
   y^6}{131072} \\ \nonumber 
   & \hspace{1.3cm} +\left(-\frac{y^2}{8}+\frac{21 y^3}{64}+\frac{709
   y^4}{512}+\frac{27597 y^5}{8192}+\frac{814491 y^6}{131072}\right) \cos
   2 \chi \bigg] \\ \nonumber 
   & + e^4 \bigg[y-\frac{11 y^2}{2}-\frac{7377 y^3}{512}-\frac{95161
   y^4}{2048}-\frac{14850039 y^5}{65536}-\frac{372021671
   y^6}{262144}\\ \nonumber
   & \hspace{0.5cm} +\left(-\frac{3 y^2}{4}+\frac{27
   y^3}{128}+\frac{3001 y^4}{512}+\frac{494913 y^5}{16384}+\frac{4559499
   y^6}{32768}\right) \cos 2 \chi \\
   & \hspace{0.5cm} +\left(\frac{45 y^3}{512}+\frac{205
   y^4}{2048}-\frac{19305 y^5}{65536}-\frac{510165 y^6}{262144}\right)
   \cos 4 \chi \bigg]\,.\label{res:Bterm_field}
\end{align}
The latter result has been computed by taking the  PN solution of the Klein-Gordon equation for $l\rightarrow+\infty$. Besides, it is in agreement with the general expression for the $B$ term computed in \cite{Heffernan:2012su}.
At this point, the explicit expression for the $\chi$-domain regularized scalar field can also be provided, using the MST solutions up to $l = 4$ (included)
\begin{align} \label{res:reg_field}\nonumber
    \psi_{\text{R}}(y,e; \chi) &= -y^3-\frac{38}{45} \pi  y^{11/2}+y^4 \left(\frac{35}{18}-\frac{4 \gamma_E
   }{3}-\frac{7 \pi ^2}{32}-\frac{4 \log 2}{3}-\frac{2 \log
   (y)}{3}\right) \\ \nonumber 
   & +y^5 \left(\frac{1141}{360}+\frac{2 \gamma_E }{3}+\frac{29
   \pi ^2}{512}-\frac{18 \log 2}{5}+\frac{\log y}{3}\right) \\ \nonumber 
   & +y^6
   \left(-\frac{23741}{1680}+\frac{77 \gamma_E }{6}-\frac{279 \pi
   ^2}{1024}+\frac{1627 \log 2}{42}-\frac{729 \log 3}{70}+\frac{77
   \log y}{12}\right) \\ \nonumber
   & + e \Bigg\{ \cos \chi  \bigg[-2 y^3-\frac{76}{9} \pi  y^{11/2}+y^4
   \left(-\frac{13}{9}-\frac{8 \gamma_E }{3}-\frac{5 \pi ^2}{32}-\frac{40
   \log 2}{3}-\frac{4 \log y}{3}\right) \\ \nonumber 
   & \hspace{1.5cm} +y^5
   \left(\frac{1031}{60}+12 \gamma_E -\frac{2709 \pi ^2}{512}+\frac{516
   \log 2}{5}-\frac{243 \log 3}{5}-2 \log y\right) \\ \nonumber 
   & \hspace{1.5cm} +y^6
   \left(-\frac{13035}{56}+\frac{29 \gamma_E }{3}+\frac{1473 \pi
   ^2}{512}-\frac{20227 \log 2}{35}+\frac{36693 \log 3}{70}+\frac{341
   \log y}{6}\right)\bigg]\\ \nonumber 
   & \hspace{0.5cm}+\bigg[-y^{5/2}+\frac{3 y^{7/2}}{2}+\frac{2
   \pi  y^4}{3}-\frac{65 y^{9/2}}{24}-\frac{146 \pi  y^5}{15}-\frac{715
   \pi  y^6}{84} \\ \nonumber
   & \hspace{0.8cm}+y^{11/2} \left(\frac{354839}{3600}-\frac{76 \gamma_E
   }{5}+4 \pi ^2-\frac{452 \log 2}{9}-\frac{78 \log
   y}{5}\right)\bigg] \sin \chi  \Bigg\} \\ \nonumber 
   & + e^2 \Bigg\{ -4 y^3-\frac{646}{45} \pi  y^{11/2}+y^5 \left(\frac{2933}{45}+\frac{50
   \gamma_E }{3}-\frac{5597 \pi ^2}{1024}+\frac{2494 \log
   2}{15}-\frac{486 \log 3}{5}-\frac{11 \log y}{3}\right) \\ \nonumber 
   & \hspace{0.8cm} +y^4
   \left(\frac{101}{9}-\frac{20 \gamma_E }{3}-\frac{13 \pi ^2}{16}-\frac{68
   \log 2}{3}-\frac{10 \log y}{3}\right) \\ \nonumber 
   & \hspace{0.8cm}+y^6
   \left(-\frac{266261}{840}+\frac{809 \gamma_E }{6}+\frac{25323 \pi
   ^2}{2048}-\frac{10729 \log 2}{10}+\frac{72657 \log
   3}{70}+\frac{1793 \log y}{12}\right) \\ \nonumber 
   & \hspace{0.8cm}+\cos 2 \chi
   \bigg[y^3-\frac{1349}{90} \pi  y^{11/2}+y^5 \left(\frac{2093}{120}+38
   \gamma_E -\frac{21183 \pi ^2}{1024}-\frac{4862 \log 2}{5}+\frac{14013
   \log 3}{20}-9 \log y\right) \\ \nonumber 
   & \hspace{1.9cm}+y^4 \left(-\frac{151}{6}+4 \gamma_E
   +\frac{3 \pi ^2}{2}+\frac{172 \log 2}{3}-\frac{81 \log 3}{2}+2
   \log y\right) -y^6 \bigg(\frac{85703}{112} \\ \nonumber 
   & \hspace{1.9cm}+\frac{1271 \gamma_E
   }{6}-\frac{65457 \pi ^2}{2048}-\frac{2070871 \log
   2}{210}+\frac{30564 \log 3}{7}+\frac{390625 \log
   5}{336}-\frac{1825 \log y}{12}\bigg)\bigg]\\ \nonumber 
   & \hspace{0.8cm}+\sin 2 \chi \bigg[-y^{5/2}+4
   y^{7/2}-\frac{11 \pi  y^4}{6}-\frac{283 y^{9/2}}{24}-\frac{241 \pi 
   y^5}{12}+\frac{6337 \pi  y^6}{70} \\ \nonumber 
   & \hspace{1.9cm}+y^{11/2}
   \left(\frac{190159}{1800}-\frac{152 \gamma_E }{5}+8 \pi ^2+\frac{2192
   \log 2}{45}-\frac{1539 \log 3}{10}-\frac{356 \log
   y}{5}\right)\bigg]  \Bigg\} \\ \nonumber 
   & + e^3 \Bigg\{ \cos 3 \chi  \bigg[\frac{3 y^3}{2}-\frac{893}{108} \pi  y^{11/2}+y^4
   \left(-\dfrac{1231}{36} + \dfrac{20\gamma_E}{3}+\dfrac{31\pi^2}{16}-\frac{2860 \log 2}{9}+\frac{837 \log 3}{4}+\frac{10 \log
   y}{3}\right) \\ \nonumber 
   & \hspace{1.8cm}+y^5
   \left(\frac{54349}{720}+\frac{80 \gamma_E }{3}-\frac{62041 \pi
   ^2}{2048}+\frac{331304 \log 2}{45}-\frac{14769 \log
   3}{5}-\frac{78125 \log 5}{72}-\frac{68 \log y}{3}\right)\\ \nonumber 
   & \hspace{1.8cm}-y^6 \bigg(\frac{246899}{288}+\frac{2527 \gamma_E
   }{6}-\frac{392859 \pi ^2}{4096}+\frac{15316951 \log
   2}{210}-\frac{53091 \log 3}{7}-\frac{52496875 \log
   5}{2016} \\ \nonumber 
   & \hspace{1.8cm}-\frac{3689 \log y}{12}\bigg)\bigg]+\cos \chi 
   \bigg[-\frac{11 y^3}{2} -\frac{11989}{180} \pi 
   y^{11/2}+y^4
   \bigg(-\frac{649}{36}-\frac{20 \gamma_E }{3}+\frac{7 \pi
   ^2}{8}+\frac{292 \log 2}{3}\\ \nonumber 
   & \hspace{1.8cm}-\frac{405 \log 3}{4}-\frac{10 \log
   y}{3}\bigg) +y^5 \bigg(\frac{156571}{720}+\frac{304 \gamma_E
   }{3}-\frac{111599 \pi ^2}{2048}-\frac{38144 \log 2}{15}+\frac{14499
   \log 3}{8}\\ \nonumber 
   & \hspace{1.8cm}-\frac{100 \log y}{3}\bigg)+y^6 \bigg(-\frac{2511063}{1120}-\frac{1057 \gamma_E
   }{6}+\frac{567333 \pi ^2}{4096}+\frac{1241395 \log
   2}{42}-\frac{6311169 \log 3}{560}\\ \nonumber 
   & \hspace{1.8cm}-\frac{390625 \log
   5}{96}+\frac{8807 \log y}{12}\bigg)\bigg]+\sin \chi
   \bigg[-\frac{11 y^{5/2}}{4}+\frac{97 y^{7/2}}{8}-\frac{5 \pi 
   y^4}{12}-\frac{565 y^{9/2}}{32}-\frac{2481 \pi  y^5}{40} \\ \nonumber 
   & \hspace{1.8cm}+\frac{39017
   \pi  y^6}{280}+y^{11/2} \left(\frac{9289201}{14400}-\frac{551 \gamma_E
   }{5}+29 \pi ^2+\frac{2803 \log 2}{45}-\frac{1539 \log
   3}{4}-\frac{1731 \log y}{10}\right)\bigg] \\ \nonumber 
   & \hspace{0.8cm}+\sin 3 \chi \bigg[-\frac{y^{5/2}}{4}+\frac{35 y^{7/2}}{8}-\frac{121 \pi 
   y^4}{36}-\frac{1621 y^{9/2}}{96}-\frac{377 \pi  y^5}{30}+\frac{1068661
   \pi  y^6}{5040} \\ \nonumber 
   & \hspace{1.8cm}+y^{11/2} \left(-\frac{709127}{4800}-\frac{247 \gamma_E
   }{15}+\frac{13 \pi ^2}{3}-\frac{23053 \log 2}{15}+\frac{15903 \log
   3}{20}-\frac{3907 \log y}{30}\right)\bigg]   \Bigg\} \\ \nonumber 
   & + e^4 \Bigg\{ -\frac{71 y^3}{8} -\frac{10697}{144} \pi  y^{11/2}+y^4
   \left(\frac{4703}{144}-\frac{103 \gamma_E }{6}-\frac{379 \pi
   ^2}{256}+\frac{329 \log 2}{6}-\frac{1377 \log 3}{16}-\frac{103
   \log y}{12}\right) \\ \nonumber 
   & \hspace{0.8cm}+y^5
   \left(\frac{63521}{192}+\frac{357 \gamma_E }{4}-\frac{161847 \pi
   ^2}{4096}-\frac{50399 \log 2}{20}+\frac{276129 \log
   3}{160}-\frac{267 \log y}{8}\right)+y^6 \bigg(-\frac{31438061}{13440} \\ \nonumber 
   & \hspace{0.8cm}+\frac{22507
   \gamma_E }{48}+\frac{1452237 \pi ^2}{8192}+\frac{57264337 \log
   2}{1680}-\frac{6320889 \log 3}{560}-\frac{14453125 \log
   5}{2688}+\frac{101227 \log y}{96}\bigg) \\ \nonumber 
   & \hspace{0.8cm}+\cos 4 \chi 
   \bigg[\frac{3 y^3}{8}+\frac{2185}{864} \pi  y^{11/2} +y^4 \bigg(-\frac{2731}{144}+\frac{19 \gamma_E
   }{6}+\frac{223 \pi ^2}{256}+\frac{25961 \log 2}{18}-\frac{16119 \log
   3}{32}\\ \nonumber 
   & \hspace{2cm}-\frac{78125 \log 5}{288} +\frac{19 \log y}{12}\bigg)  
   +y^5
   \bigg(\frac{50149}{320}-\frac{73 \gamma_E }{4}-\frac{93345 \pi
   ^2}{4096}-\frac{6150733 \log 2}{180}+\frac{1172799 \log
   3}{320}\\ \nonumber 
   & \hspace{2cm}+\frac{7035625 \log 5}{576}-\frac{249 \log
   y}{8}\bigg)+y^6
   \bigg(-\frac{12447469}{40320}-\frac{15047 \gamma_E }{48}+\frac{1135995
   \pi ^2}{8192}+\frac{363730429 \log 2}{1008}\\ \nonumber 
   & \hspace{2cm}+\frac{49057083 \log
   3}{560}-\frac{749322125 \log 5}{4032}-\frac{282475249 \log
   7}{11520}+\frac{39289 \log y}{96}\bigg)\bigg]\\ \nonumber 
   & \hspace{0.8cm}+\cos 2 \chi 
   \bigg[\frac{7 y^3}{2}-\frac{15637}{180} \pi  y^{11/2}+y^4
   \left(-\frac{4867}{36}+\frac{62 \gamma_E }{3}+\frac{469 \pi
   ^2}{64}-\frac{8950 \log 2}{9}+\frac{2511 \log 3}{4}+\frac{31 \log
   y}{3}\right)\\ \nonumber 
   & \hspace{2cm}+y^5
   \bigg(\frac{75947}{240}+171 \gamma_E -\frac{37323 \pi
   ^2}{256}+\frac{1171843 \log 2}{45}-\frac{380781 \log
   3}{40} -\frac{78125 \log 5}{18}-\frac{189 \log y}{2}\bigg)\\ \nonumber 
   & \hspace{2cm}+y^6 \bigg(-\frac{6413707}{1440}-\frac{26419 \gamma_E
   }{12}+\frac{251931 \pi ^2}{512}-\frac{50395357 \log
   2}{180}+\frac{5054157 \log 3}{280}+\frac{72521875 \log
   5}{672} \\ \nonumber 
   & \hspace{2cm}+\frac{33917 \log y}{24}\bigg)\bigg]+\sin 2 \chi \bigg[-\frac{5 y^{5/2}}{2}+\frac{83 y^{7/2}}{4}-\frac{353 \pi 
   y^4}{36}-\frac{3601 y^{9/2}}{48}-\frac{11191 \pi 
   y^5}{120}+\frac{24571 \pi  y^6}{24} \\ \nonumber 
   & \hspace{2cm}+y^{11/2}
   \left(\frac{547051}{1440}-\frac{532 \gamma_E }{3}+\frac{140 \pi
   ^2}{3}-\frac{47080 \log 2}{9}+\frac{47709 \log 3}{20}-\frac{1646
   \log y}{3}\right)\bigg] \\ \nonumber 
   & \hspace{0.8cm}+\sin 4 \chi \bigg[\frac{17
   y^{7/2}}{8}-\frac{457 \pi  y^4}{288}-\frac{611 y^{9/2}}{48}+\frac{8491
   \pi  y^5}{960}+\frac{851239 \pi  y^6}{5376}+y^{11/2}
   \bigg(-\frac{4896077}{14400}\\  
   & \hspace{2cm}+\frac{76 \gamma_E }{15}-\frac{4 \pi
   ^2}{3}+\frac{923608 \log 2}{135}-\frac{306261 \log
   3}{160}-\frac{1484375 \log 5}{864}-\frac{1822 \log
   y}{15}\bigg)\bigg]  \Bigg\} \,, 
\end{align}
\end{widetext}
where $\gamma_E$ is the Euler-Mascheroni constant, and the result reduces to the circular case in the limit $e \to 0$. \textcolor{black}{In} Fig.~\ref{fig:Fig_psiR}, we show the impact of the eccentricity with respect to $y$ for different values of $e$ and different angular configurations, i.e. $\chi = 0,  \pi, 2\pi$. By carefully looking at Eq.~\eqref{res:reg_field}, it is clear that the only $\chi$-dependence of the field is through trigonometric functions, that reflect the periodic motion of the source. This is also shown in the aforementioned plot, where the two lines for $\chi=0$ and $2\pi$ are completely overlapping.

In the next Section, we will use the results found above to compute the Self-Force components.

\section{Scalar Self-Force}\label{sec_4}

We recall that the unit-mass SSF acting on a  scalar charge is given by
\begin{equation}
    F_\alpha(t) = \dfrac{q}{\mu}P^\beta_\alpha (t)\nabla_\beta \psi_{\text{R}},
\end{equation}
where $\psi_R$ is the regularized field (discussed in the previous Section) evaluated at the particle position and $P^\beta_\alpha (t)=\delta_\alpha^\beta + u_\alpha u^\beta$ projects orthogonally to the particle’s world-line. It is important to point out that the scalar force scales as $F_\alpha(t)\propto q^2/\mu M^2$, or alternatively $F_\alpha(t)=O(\varepsilon/M)$. By considering this force small, we are indeed asking that   
\begin{equation}\label{eq:varepsilon}
    \varepsilon = \dfrac{q^2}{\mu M} \ll 1\,.
\end{equation}
These considerations will be important in later stages of this paper (Sec.~\ref{sec_6}), where we will compare our results with independent computations performed in scalar-tensor theories with the post-Newtonian formalism.

Since we are restricting the motion to the equatorial plane, there are three non-vanishing components of the self-force which take the following form
\begin{align}
    F_t (t) &= \dfrac{q}{\mu}\left(1-u_t^2 f(r_p(t))\right)\partial_t \psi_{\rm R} \nonumber\\
    &\quad -\dfrac{q}{\mu}f(r_p(t))u_t\left(  u_r  \partial_r \psi_{\rm R} +u_{\phi }\partial_\phi \psi_{\rm R} \right) , \label{Ft formula}\\
    F_r (t) &= \dfrac{q}{\mu}(1+u_r^2 f^{-1}(r_p(t)))\partial_r\psi_{\rm R}\nonumber\\
    &\quad+\dfrac{q}{\mu}u_r f^{-1}(r_p(t)) \left( u_t \partial_t \psi_{\rm R} + u_\phi \partial_\phi \psi_{\rm R}  \right), \\
    F_{\phi} (t) &= \dfrac{q}{\mu}(1+r^2_p(t) u_\phi^2)\partial_\phi\psi_{\rm R}\nonumber\\
    &\quad +\dfrac{q}{\mu}u_\phi r^2_p(t) \left( u_t \partial_t\psi_{\rm R} + u_r \partial_r\psi_{\rm R}\right) .\label{Fphi formula}
\end{align}

In analogy with the calculations performed in the previous Section, in order to construct the derivative of the regularized scalar field, we first have to compute the same derivatives for the retarded field, i.e.
\begin{widetext}
    \begin{equation}
    \partial_t \psi^{\rm ret}(t) = \sum_{lm} \partial_t \left( e^{-im\Omega_\phi t} \bar{\psi}_{lm}(t,r)\right) Y_{lm}(\theta,\phi)\Bigg|_{(r,\theta,\phi)\to (r_p(t),\pi/2,\phi_p(t))}\,,
\end{equation}
\begin{equation}
    \partial_r \psi^{\rm ret}(t) =  \sum_{lm} e^{-im\Omega_\phi t} \partial_r\bar{\psi}_{lm}(t,r) Y_{lm}(\theta,\phi)\Bigg|_{(r,\theta,\phi)\to (r_p(t),\pi/2,\phi_p(t))}\,,
\end{equation}
\begin{equation}
    \partial_\phi \psi^{\rm ret}(t) = \sum_{lm} (im) e^{-im\Omega_\phi t} \bar{\psi}_{lm}(t,r) Y_{lm}(\theta,\phi)\Bigg|_{(r,\theta,\phi)\to (r_p(t),\pi/2,\phi_p(t))}\,,
\end{equation}
\end{widetext}

which needs to be regularized. Alternatively, we can directly compute the non-regularized components of the force with the derivatives of the scalar field and then use the mode-sum regularization directly on the forces. For our purposes, we will follow the first methodology but we checked that the two methods are indeed equivalent.

Accordingly, the regularized self-force is obtained by subtracting the divergent term $B_\alpha(t)$ after performing the $m$-mode summation and evaluating at the particle’s position as follows
\begin{equation}
    F_\alpha^R(t) = \sum_l \left\{ \dfrac{1}{2}\left(F_{\alpha\,(+)}^l(t) - F_{\alpha\,(-)}^l(t)\right) - B_\alpha(t) \right\},
\end{equation}
where the $\pm$ is used to show that we are evaluating the $l$-modes of the force components at the particle position $r\rightarrow r_p^{\pm}(t)$. 

By using the same convention as before, the subtraction terms are (in units of $\varepsilon$) as functions of $\chi$
\begin{widetext}
\begin{subequations}
        \begin{align}\nonumber
        B_t(y,e;\chi) &= e \sin \chi \left(\frac{y^{5/2}}{2}-\frac{5 y^{7/2}}{8}-\frac{375
   y^{9/2}}{128}-\frac{6181 y^{11/2}}{512}\right) + e^2 \sin 2 \chi \left(\frac{y^{5/2}}{2}-\frac{y^{7/2}}{2}-\frac{749
   y^{9/2}}{128}-\frac{5531 y^{11/2}}{256}\right) \\ \nonumber
   & \hspace{-0.5cm} + e^3 \bigg[\left(\frac{11 y^{5/2}}{8}-\frac{33 y^{7/2}}{8}-\frac{10983
   y^{9/2}}{512}-\frac{75889 y^{11/2}}{1024}\right) \sin \chi
   +\left(\frac{y^{5/2}}{8}+\frac{y^{7/2}}{16}-\frac{2761
   y^{9/2}}{512}-\frac{6933 y^{11/2}}{512}\right) \sin 3 \chi \bigg] \\ 
   & \hspace{-0.5cm} + e^4 \bigg[\left(\frac{5 y^{5/2}}{4}-\frac{7 y^{7/2}}{2}-\frac{9131
   y^{9/2}}{256}-\frac{24751 y^{11/2}}{256}\right) \sin 2 \chi
   +\left(\frac{y^{7/2}}{8}-\frac{779 y^{9/2}}{256}-\frac{845
   y^{11/2}}{1024}\right) \sin 4 \chi \bigg] \,,\\ \nonumber
        B_r(y,e;\chi) &= -\frac{y^2}{2}-\frac{y^3}{8}-\frac{21 y^4}{128}-\frac{53
   y^5}{512}+\frac{12607 y^6}{32768} + e \cos \chi \left(-y^2-\frac{y^3}{2}-\frac{15 y^4}{64}+\frac{73 y^5}{64}+\frac{93563
   y^6}{16384}\right) \\ \nonumber 
   & \hspace{-0.5cm} + e^2 \bigg[ -\frac{5 y^2}{4}+y^3+\frac{1661 y^4}{256}+\frac{16547
   y^5}{512}+\frac{10397703
   y^6}{65536}+\cos 2
   \chi\left(-\frac{y^2}{4}-\frac{y^3}{8}+\frac{101
   y^4}{256}-\frac{13 y^5}{64}-\frac{475373 y^6}{65536}\right)   \bigg] \\ \nonumber 
   & \hspace{-0.5cm} + e^3 \bigg[ \cos \chi\left(-2 y^2+\frac{15 y^3}{8}+\frac{1981 y^4}{128}+\frac{4739
   y^5}{64}+\frac{5584373 y^6}{16384}\right) 
   +\cos 3 \chi\left(\frac{y^3}{8}+\frac{103 y^4}{128}-\frac{179
   y^5}{64}-\frac{396111 y^6}{16384}\right)   \bigg] \\ \nonumber 
   & \hspace{-0.5cm} + e^4 \bigg[ -2 y^2+\frac{273 y^3}{64}+\frac{27663 y^4}{1024}+\frac{269205
   y^5}{2048}+\frac{83361301 y^6}{131072} \\ 
   & \hspace{-0.1cm} -\cos 2 \chi\left(\frac{y^2}{2}-\frac{9
   y^3}{16}-\frac{1689 y^4}{256}-\frac{6273 y^5}{512}-\frac{34369
   y^6}{2048}\right)  +\cos 4 \chi\left(\frac{3 y^3}{64}+\frac{581
   y^4}{1024}-\frac{6157 y^5}{2048}-\frac{2618153 y^6}{131072}\right)
     \bigg] \,,\\ \nonumber
        B_\phi(y,e;\chi) &= e \sin \chi \left(-\frac{3 y^2}{4} +\frac{15 y^3}{32}+\frac{1119
   y^4}{256}+\frac{83079 y^5}{4096}+\frac{5078371 y^6}{65536}\right) \\ \nonumber
   & \hspace{-0.5cm}+ e^2 \left(-\frac{3 y^2}{4}+\frac{21 y^3}{16}+\frac{1785
   y^4}{256}+\frac{29217 y^5}{1024}+\frac{8221319 y^6}{65536}\right)
   \sin 2 \chi \\ \nonumber 
   & \hspace{-0.5cm} + e^3 \bigg[  \sin \chi  \left(-\frac{27 y^2}{16}+\frac{303 y^3}{64}+\frac{26265
   y^4}{1024}+\frac{108537 y^5}{1024}+\frac{127678143
   y^6}{262144}\right) \\ \nonumber 
   & \hspace{-0.1cm}+\sin 3 \chi\left(-\frac{3
   y^2}{16} +\frac{69 y^3}{64}+\frac{3369 y^4}{1024}+\frac{44913
   y^5}{4096}+\frac{15633911 y^6}{262144}\right)   \bigg] \\ \nonumber
   & \hspace{-0.5cm} + e^4 \bigg[ \sin 2 \chi \left(-\frac{3 y^2}{2} +\frac{219 y^3}{32}+\frac{7419
   y^4}{256}+\frac{214545 y^5}{2048}+\frac{8669885 y^6}{16384}\right)
     \\ 
     & \hspace{-0.1cm}+\sin 4 \chi \left(\frac{21 y^3}{64}-\frac{9 y^4}{512}-\frac{3249
   y^5}{4096}-\frac{534939 y^6}{65536}\right) \bigg] \,.
    \end{align}
\end{subequations}

We can ultimately show the final result for the three SSF components, reading as
\begin{subequations}
        \begin{align} \nonumber
        F_t(y,e;\chi) &= \frac{y^4}{3}-\frac{y^5}{6}+\frac{2}{3} \pi  y^{11/2}-\frac{77 y^6}{24} + e \Bigg\{ \left(\frac{5 y^4}{3}-\frac{5 y^5}{2}+6 \pi  y^{11/2}-\frac{653
   y^6}{24}\right) \cos \chi  \\ \nonumber 
   & +\left[2 y^{9/2}+y^{11/2}
   \left(-\frac{187}{9}+\frac{28 \gamma_E }{3}+\frac{35 \pi ^2}{64}+20
   \log 2+\frac{14 \log y}{3}\right)\right] \sin \chi  \Bigg\} \\ \nonumber 
   & + e^2 \Bigg\{ \frac{8 y^4}{3}-\frac{13 y^5}{6}+\frac{32}{3} \pi  y^{11/2}-\frac{299
   y^6}{4}+\cos 2 \chi\left(2 y^4-\frac{23 y^5}{2}+\frac{89}{6} \pi 
   y^{11/2}-\frac{91 y^6}{2}\right) \\ \nonumber 
   & \hspace{0.5cm}+ \sin 2 \chi \left[5
   y^{9/2}+y^{11/2} \left(-\frac{1343}{18}+\frac{88 \gamma_E
   }{3}+\frac{119 \pi ^2}{64}+\frac{56 \log 2}{3}+\frac{81 \log
   3}{2}+\frac{44 \log y}{3}\right)\right]   \Bigg\} \\ \nonumber 
   & + e^3 \Bigg\{ \cos \chi \left(\frac{26 y^4}{3}-\frac{82 y^5}{3}+\frac{331}{6} \pi 
   y^{11/2}-\frac{1245 y^6}{4}\right)  +\cos 3 \chi\left(\frac{4
   y^4}{3}-\frac{53 y^5}{3}+\frac{181}{9} \pi  y^{11/2}-\frac{152
   y^6}{3}\right) \\ \nonumber 
   & \hspace{0.5cm} +\sin \chi \left[14 y^{9/2}+y^{11/2}
   \left(-\frac{4105}{18}+\frac{251 \gamma_E }{3}+\frac{1309 \pi
   ^2}{256}+\frac{139 \log 2}{3}+\frac{243 \log 3}{2}+\frac{251 \log
   y}{6}\right)\right] \\ \nonumber 
   & \hspace{0.5cm} +\sin 3 \chi \left[5 y^{9/2}+y^{11/2}
   \left(-\frac{1051}{9}+\frac{121 \gamma_E }{3}+\frac{707 \pi
   ^2}{256}+275 \log 2-81 \log 3+\frac{121 \log
   y}{6}\right)\right]  \Bigg\} + \\ \nonumber 
   & + e^4 \Bigg\{ \frac{217 y^4}{24}-\frac{275 y^5}{16}+\frac{2831}{48} \pi 
   y^{11/2}-\frac{31703 y^6}{64}+ \cos 2 \chi\left(\frac{53 y^4}{6}-\frac{333
   y^5}{4}+\frac{3781}{36} \pi  y^{11/2}-\frac{15301 y^6}{48}\right)\\ \nonumber 
   & \hspace{0.5cm} +\cos 4 \chi \left(\frac{11 y^4}{24}-\frac{193
   y^5}{16}+\frac{4421}{288} \pi  y^{11/2}-\frac{9823 y^6}{192}\right)
    \\ \nonumber 
    & \hspace{0.5cm} +\sin 2 \chi \left[\frac{55 y^{9/2}}{2}+y^{11/2}
   \left(-\frac{2483}{4}+\frac{620 \gamma_E }{3}+\frac{1715 \pi
   ^2}{128}+\frac{10180 \log 2}{9}-\frac{1053 \log 3}{4}+\frac{310
   \log y}{3}\right)\right] \\
   & \hspace{0.5cm} + \sin 4 \chi \left[\frac{5
   y^{9/2}}{2}-y^{11/2} \left(\frac{895}{9}-\frac{92 \gamma_E
   }{3}-\frac{287 \pi ^2}{128}+\frac{1964 \log 2}{3}-\frac{3321 \log
   3}{32}-\frac{78125 \log 5}{288}-\frac{46 \log
   y}{3}\right)\right] \Bigg\} \,,\\ \nonumber
        F_r(y,e;\chi) &= y^5 \left(-\frac{2}{9}-\frac{4 \gamma_E }{3}+\frac{7 \pi ^2}{64}-\frac{4
   \log 2}{3}-\frac{2 \log y}{3}\right)+y^6
   \left(\frac{604}{45}-\frac{14 \gamma_E }{3}+\frac{29 \pi
   ^2}{1024}-\frac{66 \log 2}{5}-\frac{7 \log y}{3}\right) \\ \nonumber 
   & + e \Bigg\{ \cos \chi  \bigg[y^5 \left(\frac{92}{9}-\frac{32 \gamma_E }{3}+\frac{7
   \pi ^2}{32}-\frac{64 \log 2}{3}-\frac{16 \log y}{3}\right) \\ \nonumber 
   & \hspace{1.5cm}+y^6
   \left(\frac{3694}{45}-\frac{8 \gamma_E }{3}+\frac{565 \pi
   ^2}{128}+\frac{512 \log 2}{5}-\frac{486 \log 3}{5}+\frac{20 \log
   y}{3}\right)\bigg] \\ \nonumber 
   & \hspace{0.5cm}+\sin \chi\left(\frac{2 y^{7/2}}{3}-8 y^{9/2}+4 \pi 
   y^5-\frac{61 y^{11/2}}{4}-\frac{154 \pi  y^6}{15}\right)   \Bigg\} \\ \nonumber 
   & + e^2 \Bigg\{ y^5 \left(\frac{136}{9}-\frac{58 \gamma_E }{3}+\frac{77 \pi
   ^2}{128}-\frac{122 \log 2}{3}-\frac{29 \log y}{3}\right) \\ \nonumber 
   & \hspace{0.8cm}+y^6
   \left(\frac{10136}{45}-\frac{67 \gamma_E }{3}+\frac{1543 \pi
   ^2}{256}+\frac{653 \log 2}{3}-243 \log 3+\frac{29 \log
   y}{6}\right) \\ \nonumber 
   & \hspace{0.8cm}+\cos 2 \chi  \bigg[y^5 \left(\frac{368}{9}-\frac{62
   \gamma_E }{3}-\frac{77 \pi ^2}{128}+\frac{2 \log 2}{3}-\frac{81 \log
   3}{2}-\frac{31 \log y}{3}\right) \\ \nonumber 
   & \hspace{2cm}+y^6
   \left(\frac{4142}{45}+\frac{227 \gamma_E }{3}+\frac{8543 \pi
   ^2}{512}-1247 \log 2+\frac{17577 \log 3}{20}+\frac{419 \log
   y}{6}\right)\bigg] \\ \nonumber 
   & \hspace{0.8cm}+ \sin 2 \chi \left(y^{7/2}-\frac{44 y^{9/2}}{3}+\frac{61 \pi
    y^5}{6}-\frac{1357 y^{11/2}}{24}-\frac{2641 \pi  y^6}{60}\right)
   \Bigg\} \\ \nonumber 
   & + e^3 \Bigg\{ \cos 3 \chi  \bigg[y^5 \left(\frac{991}{18}-\frac{64 \gamma_E
   }{3}-\frac{77 \pi ^2}{64}-\frac{2512 \log 2}{9}+\frac{243 \log
   3}{2}-\frac{32 \log y}{3}\right) \\ \nonumber 
   & \hspace{1.8cm}+y^6
   \left(\frac{3679}{36}+\frac{418 \gamma_E }{3}+\frac{11347 \pi
   ^2}{512}+\frac{442168 \log 2}{45}-\frac{54351 \log
   3}{20}-\frac{78125 \log 5}{36}+\frac{359 \log
   y}{3}\right)\bigg] \\ \nonumber 
   & \hspace{0.8cm}+\cos \chi  \bigg[y^5
   \left(\frac{2213}{18}-\frac{256 \gamma_E }{3}+\frac{7 \pi ^2}{64}-48
   \log 2-\frac{243 \log 3}{2}-\frac{128 \log y}{3}\right) \\ \nonumber 
   & \hspace{1.8cm}+y^6
   \left(\frac{14753}{20}+154 \gamma_E +\frac{27057 \pi
   ^2}{512}-\frac{60448 \log 2}{15}+\frac{25353 \log 3}{10}+195 \log
   y\right)\bigg] \\ \nonumber 
   & \hspace{0.8cm}+\sin \chi
   \left(\frac{17 y^{7/2}}{6}-\frac{201
   y^{9/2}}{4}+\frac{181 \pi  y^5}{6}-\frac{5257
   y^{11/2}}{48}-\frac{4513 \pi  y^6}{30}\right) \\ \nonumber 
   & \hspace{0.8cm}+\sin 3\chi
   \left(\frac{y^{7/2}}{2}-\frac{37 y^{9/2}}{4}+\frac{193 \pi 
   y^5}{18}-\frac{1429 y^{11/2}}{16}-\frac{4157 \pi  y^6}{60}\right) \Bigg\} \\ \nonumber 
   & + e^4 \Bigg\{ y^5 \left(\frac{8363}{72}-\frac{569 \gamma_E }{6}+\frac{707 \pi
   ^2}{512}-\frac{313 \log 2}{6}-\frac{2187 \log 3}{16}-\frac{569
   \log y}{12}\right) \\ \nonumber 
   & \hspace{0.8cm} +y^6 \left(\frac{96289}{80}+\frac{323 \gamma_E
   }{4}+\frac{199149 \pi ^2}{4096}-\frac{332851 \log
   2}{60}+\frac{517347 \log 3}{160}+\frac{1411 \log
   y}{8}\right) \\ \nonumber 
   & \hspace{0.8cm}+\cos 4 \chi \bigg[y^5
   \left(\frac{2425}{72}-\frac{71 \gamma_E }{6}-\frac{427 \pi
   ^2}{512}+\frac{16811 \log 2}{18}-\frac{7371 \log
   3}{32}-\frac{78125 \log 5}{288}-\frac{71 \log y}{12}\right) \\ \nonumber
   & \hspace{1.8cm}+y^6
   \left(\frac{25087}{144}+\frac{373 \gamma_E }{4}+\frac{48147 \pi
   ^2}{4096}-\frac{817943 \log 2}{20}-\frac{84969 \log
   3}{64}+\frac{10785625 \log 5}{576}+\frac{693 \log
   y}{8}\right)\bigg]\\ \nonumber 
   & \hspace{0.8cm}+\cos 2 \chi \bigg[y^5
   \left(\frac{1711}{6}-\frac{380 \gamma_E }{3}-\frac{35 \pi
   ^2}{8}-\frac{10036 \log 2}{9}+405 \log 3-\frac{190 \log
   y}{3}\right) \\ \nonumber 
   & \hspace{1.8cm}+y^6 \left(\frac{76943}{180}+776 \gamma_E +\frac{135639
   \pi ^2}{1024}+\frac{1797784 \log 2}{45}-\frac{379323 \log
   3}{40}-\frac{78125 \log 5}{8}+660 \log
   y\right)\bigg] \\ \nonumber 
   & \hspace{0.8cm}+\sin 2 \chi\left(\frac{11 y^{7/2}}{3}-\frac{309
   y^{9/2}}{4}+\frac{185 \pi  y^5}{3}-\frac{4051
   y^{11/2}}{12}-\frac{52057 \pi  y^6}{120}\right) \\  
   & \hspace{0.8cm}+ \sin 4 \chi \left(\frac{y^{7/2}}{12}-\frac{15 y^{9/2}}{8}+\frac{1699 \pi 
   y^5}{288}-\frac{2519 y^{11/2}}{32}-\frac{44617 \pi  y^6}{960}\right)
   \Bigg\} \,,\\ \nonumber
        F_\phi(y,e;\chi) &= -\frac{y^{5/2}}{3}+\frac{y^{7/2}}{6}-\frac{2 \pi  y^4}{3}+\frac{77
   y^{9/2}}{24}-\frac{9 \pi  y^5}{5}+\frac{601 \pi  y^6}{84}-y^{11/2}
   \left(\frac{10121}{3600}-\frac{76 \gamma_E }{45}+\frac{4 \pi
   ^2}{9}-\frac{76 \log 2}{45}-\frac{38 \log y}{45}\right) \\ \nonumber 
   & + e \Bigg\{ \cos \chi  \bigg[-y^{5/2}+\frac{3 y^{7/2}}{2}-\frac{14 \pi 
   y^4}{3}+\frac{475 y^{9/2}}{24}+\frac{21 \pi  y^5}{5}+\frac{53167 \pi 
   y^6}{420} \\ \nonumber 
   & \hspace{1.8cm}+y^{11/2} \left(-\frac{98161}{720}+\frac{76 \gamma_E
   }{3}-\frac{20 \pi ^2}{3}+\frac{2356 \log 2}{45}+\frac{38 \log
   y}{3}\right)\bigg] \\ \nonumber
   & \hspace{0.8cm}+\sin \chi \bigg[-2 y^3-\frac{532}{45} \pi  y^{11/2}+y^4
   \left(21-8 \gamma_E -\frac{21 \pi ^2}{32}-\frac{56 \log 2}{3}-4 \log
   y\right) \\ \nonumber 
   & \hspace{1.8cm}+y^5 \left(\frac{7981}{180}+\frac{76 \gamma_E }{3}+\frac{773
   \pi ^2}{256}+\frac{804 \log 2}{5}-\frac{486 \log 3}{5}+\frac{38
   \log y}{3}\right) \\ \nonumber 
   & \hspace{1.8cm}+y^6 \left(-\frac{10565}{24}+\frac{929 \gamma_E
   }{3}+\frac{2193 \pi ^2}{1024}-\frac{31723 \log 2}{35}+\frac{7533
   \log 3}{7}+\frac{929 \log y}{6}\right)\bigg] \Bigg\} \\ \nonumber 
   & + e^2 \Bigg\{ -\frac{4 y^{5/2}}{3}+\frac{8 y^{7/2}}{3}-\frac{20 \pi 
   y^4}{3}+\frac{127 y^{9/2}}{3}-\frac{44 \pi  y^5}{15}+\frac{16279 \pi 
   y^6}{70} \\ \nonumber 
   & \hspace{0.8cm}+y^{11/2} \left(-\frac{142763}{900}+\frac{1672 \gamma_E }{45}-\frac{88 \pi ^2}{9}+\frac{3496 \log 2}{45}+\frac{836 \log y}{45}\right) \\ \nonumber 
   & \hspace{0.8cm}+\cos 2 \chi  \bigg[-\frac{y^{5/2}}{2}+\frac{13
   y^{7/2}}{4}-\frac{47 \pi  y^4}{6}+\frac{1223 y^{9/2}}{48}+\frac{157
   \pi  y^5}{4}+\frac{13879 \pi  y^6}{30} \\ \nonumber 
   & \hspace{1.8cm}+y^{11/2}
   \left(-\frac{942211}{1440}+\frac{266 \gamma_E }{3}-\frac{70 \pi
   ^2}{3}+\frac{190 \log 2}{9}+\frac{1539 \log 3}{10}+\frac{133 \log
   y}{3}\right)\bigg] \\ \nonumber
   & \hspace{0.8cm}+\sin 2 \chi \bigg[-3 y^3-\frac{4009}{90} \pi  y^{11/2}+y^4
   \left(\frac{93}{2}-16 \gamma_E -\frac{21 \pi ^2}{16}+\frac{32 \log
   2}{3}-\frac{81 \log 3}{2}-8 \log y\right)\\ \nonumber 
   & \hspace{1.8cm}+y^5
   \left(\frac{24539}{360}+\frac{244 \gamma_E }{3}+\frac{1715 \pi
   ^2}{256}-\frac{20612 \log 2}{15}+\frac{19521 \log
   3}{20}+\frac{122 \log y}{3}\right) \\ \nonumber 
   & \hspace{1.8cm}-y^6
   \bigg(\frac{10832999}{5040}-\frac{2998 \gamma_E }{3}+\frac{3655 \pi
   ^2}{256}-\frac{429400 \log 2}{21}+\frac{1521369 \log
   3}{280}+\frac{390625 \log 5}{112}\\ \nonumber 
   & \hspace{2.5cm}-\frac{1511 \log
   y}{3}\bigg)\bigg]  \Bigg\} \\ \nonumber 
   & + e^3 \Bigg\{ \cos 3 \chi \bigg[-\frac{y^{5/2}}{12}+\frac{73 y^{7/2}}{24}-\frac{217
   \pi  y^4}{36}+\frac{1829 y^{9/2}}{96}+\frac{4381 \pi 
   y^5}{72}+\frac{162293 \pi  y^6}{240} \\ \nonumber 
   & \hspace{1.8cm}+y^{11/2}
   \left(-\frac{10109743}{8640}+\frac{1273 \gamma_E }{9}-\frac{335 \pi
   ^2}{9}+\frac{189031 \log 2}{135}-\frac{10773 \log
   3}{20}+\frac{1273 \log y}{18}\right)\bigg] \\ \nonumber 
   & \hspace{0.8cm}+\cos \chi 
   \bigg[-\frac{11 y^{5/2}}{4} +\frac{87
   y^{7/2}}{8}-\frac{319 \pi  y^4}{12}+\frac{4575 y^{9/2}}{32}+\frac{655
   \pi  y^5}{8}+\frac{916021 \pi  y^6}{560} \\ \nonumber 
   & \hspace{1.8cm}+y^{11/2}
   \left(-\frac{4967561}{2880}+\frac{779 \gamma_E }{3}-\frac{205 \pi
   ^2}{3}+\frac{5909 \log 2}{45}+\frac{1539 \log 3}{4}+\frac{779
   \log y}{6}\right)\bigg] \\ \nonumber 
   & \hspace{0.8cm}+\sin \chi \bigg[-\frac{15
   y^3}{2} -\frac{21109}{180} \pi  y^{11/2}+y^4 \left(\frac{557}{4}-44
   \gamma_E -\frac{231 \pi ^2}{64}+\frac{52 \log 2}{3}-\frac{405 \log
   3}{4}-22 \log y\right)\\ \nonumber 
   & \hspace{1.8cm}+y^5 \left(\frac{41569}{240}+250 \gamma_E
   +\frac{25407 \pi ^2}{1024}-\frac{58534 \log 2}{15}+\frac{109431
   \log 3}{40}+125 \log y\right)\\ \nonumber
   & \hspace{1.8cm}+y^6
   \bigg(-\frac{68619637}{10080}+\frac{9892 \gamma_E }{3}-\frac{34931 \pi
   ^2}{2048}+\frac{1366471 \log 2}{21}-\frac{2215809 \log
   3}{140}-\frac{390625 \log 5}{32} \\ \nonumber 
   & \hspace{2.5cm}+\frac{4946 \log
   y}{3}\bigg)\bigg] +\sin 3 \chi \bigg[-\frac{3
   y^3}{2}-\frac{38171}{540} \pi  y^{11/2}+y^4
   \bigg(\frac{475}{12}-12 \gamma_E -\frac{63 \pi ^2}{64} \\ \nonumber 
   & \hspace{1.8cm}-\frac{844 \log
   2}{3}+\frac{567 \log 3}{4}-6 \log y\bigg)+y^5
   \bigg(-\frac{22409}{720}+\frac{358 \gamma_E }{3}+\frac{5435 \pi
   ^2}{1024}+\frac{96046 \log 2}{9} \\ \nonumber
   & \hspace{1.8cm}-\frac{131463 \log
   3}{40}-\frac{78125 \log 5}{36}+\frac{179 \log y}{3}\bigg)+y^6
   \bigg(-\frac{1197899}{288}+\frac{4030 \gamma_E }{3} \\ \nonumber 
   & \hspace{1.8cm}-\frac{130115 \pi
   ^2}{2048}-\frac{38386721 \log 2}{315}-\frac{20439 \log
   3}{7}+\frac{115853125 \log 5}{2016}+\frac{2063 \log
   y}{3}\bigg)\bigg]  \Bigg\} \\ \nonumber
   & + e^4 \Bigg\{ -\frac{65 y^{5/2}}{24} +\frac{509 y^{7/2}}{48}-\frac{391
   \pi  y^4}{16}+\frac{11571 y^{9/2}}{64}+\frac{18253 \pi 
   y^5}{480}+\frac{3933581 \pi  y^6}{2240} \\ \nonumber 
   & \hspace{0.8cm}+y^{11/2}
   \left(-\frac{13525403}{9600}+\frac{10393 \gamma_E }{45}-\frac{547 \pi
   ^2}{9}+133 \log 2+\frac{26163 \log 3}{80}+\frac{10393 \log
   y}{90}\right) \\ \nonumber
   & \hspace{0.8cm}+\cos 4 \chi  \bigg[\frac{5 y^{7/2}}{4}-\frac{577
   \pi  y^4}{288}+\frac{233 y^{9/2}}{24}+\frac{15641 \pi 
   y^5}{320}+\frac{7160983 \pi  y^6}{16128} \\ \nonumber 
   & \hspace{1.8cm}+y^{11/2}
   \bigg(-\frac{100663}{96}+\frac{703 \gamma_E }{6}-\frac{185 \pi
   ^2}{6}-\frac{1405373 \log 2}{270}+\frac{180063 \log
   3}{160}+\frac{1484375 \log 5}{864} \\ \nonumber 
   & \hspace{2.8cm}+\frac{703 \log
   y}{12}\bigg)\bigg] \\ \nonumber 
   & \hspace{0.8cm}+\cos 2 \chi  \bigg[-\frac{5
   y^{5/2}}{4} +\frac{117 y^{7/2}}{8}-\frac{1267 \pi 
   y^4}{36}+\frac{12683 y^{9/2}}{96} +\frac{11709 \pi 
   y^5}{40}+\frac{3779731 \pi  y^6}{1008}\\ \nonumber 
   & \hspace{1.8cm}+y^{11/2}
   \left(-\frac{15245411}{2880}+\frac{1957 \gamma_E }{3}-\frac{515 \pi
   ^2}{3}+\frac{44479 \log 2}{9}-\frac{32319 \log 3}{20}+\frac{1957
   \log y}{6}\right)\bigg]\\ \nonumber 
   & \hspace{0.8cm}+\sin 2 \chi \bigg[-\frac{19
   y^3}{2} -\frac{177289}{540} \pi  y^{11/2}+y^4
   \left(\frac{2939}{12}-72 \gamma_E -\frac{189 \pi ^2}{32}-\frac{2912
   \log 2}{3}+\frac{1701 \log 3}{4}-36 \log y\right) \\ \nonumber 
   & \hspace{1.8cm}+y^5
   \left(-\frac{4769}{720}+612 \gamma_E +\frac{2889 \pi
   ^2}{64}+\frac{349616 \log 2}{9}-\frac{424683 \log
   3}{40}-\frac{78125 \log 5}{9}+306 \log y\right) \\ \nonumber 
   & \hspace{1.8cm}+y^6
   \bigg(-\frac{43254473}{2016}+7974 \gamma_E -\frac{219923 \pi
   ^2}{1024}-\frac{4260481 \log 2}{9}-\frac{857007 \log
   3}{28}+\frac{161528125 \log 5}{672} \\ \nonumber 
   & \hspace{2.5cm}+4021 \log y\bigg)\bigg]\\ \nonumber 
   &\hspace{0.8cm}+\sin 4 \chi \bigg[-\frac{y^3}{4}-\frac{253061 \pi 
   y^{11/2}}{4320}+y^4 \bigg(\frac{377}{24}-4 \gamma_E -\frac{21 \pi
   ^2}{64}+\frac{9656 \log 2}{9}-\frac{9477 \log 3}{32}\\ \nonumber 
   & \hspace{1.8cm}-\frac{78125
   \log 5}{288}-2 \log y\bigg)+y^5
   \bigg(-\frac{176371}{1440}+\frac{296 \gamma_E }{3}+\frac{725 \pi
   ^2}{512}-\frac{2057038 \log 2}{45}+\frac{34101 \log
   3}{320} \\ \nonumber 
   & \hspace{1.8cm}+\frac{11410625 \log 5}{576}+\frac{148 \log
   y}{3}\bigg)+y^6 \bigg(-\frac{79561903}{20160}+882 \gamma_E
   -\frac{207065 \pi ^2}{2048}+\frac{128020547 \log
   2}{210} \\  
   & \hspace{1.8cm}+\frac{1068831207 \log 3}{4480}-\frac{1345738375 \log
   5}{4032}-\frac{282475249 \log 7}{3840}+466 \log y\bigg)\bigg]
    \Bigg\} \,.\label{eq:F_phi}
    \end{align}
\end{subequations}
\end{widetext}

\begin{figure*}[t]
  \centering
  \includegraphics[width=\linewidth]{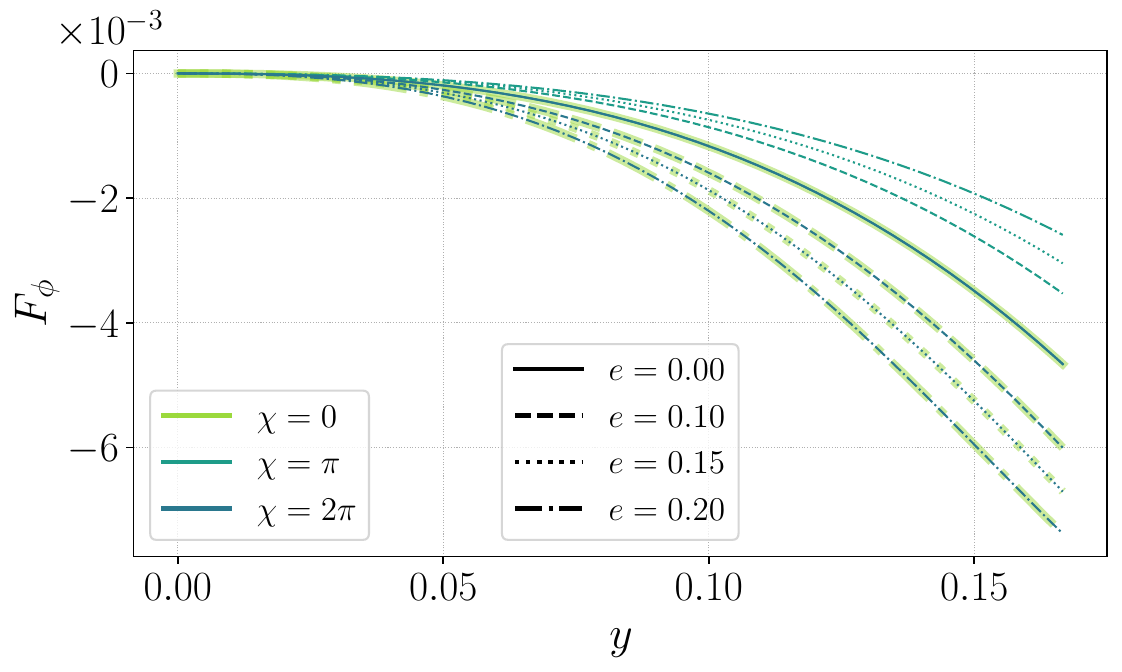}
  \caption{Here, we show the $\phi$-component of the SF. We selected the same value for both $\chi$ and $e$. The behaviour is rather similar to the one presented by the regularized field.}
  \label{fig:FPHI}
\end{figure*}

In analogy with the field, we show the $\phi$-component of the SF and its difference with respect to the circular case in Figure \ref{fig:FPHI}. 

Further, it is important to mention that the forces contain both conservative and dissipative contributions. In the next Sections we will focus on the dissipative sector by directly computing the fluxes.

\section{Scalar Radiation}\label{sec_5}

In this Section, the amount of scalar radiation emitted to infinity is computed by averaging over an orbital cycle. To this end, the asymptotic solutions of Eq.~\eqref{eq_KG_fourier} are constructed by imposing the appropriate boundary conditions, namely, ingoing waves at the horizon and purely outgoing waves at infinity. A complete derivation of the energy flux at infinity is provided below. For a more general treatment, including electromagnetic and gravitational perturbations we refer the interested reader to the original papers by Teukolsky~\cite{Teukolsky:1973ha,Press:1973zz,Teukolsky:1974yv}.

The starting point is to organize the solution of the KG equation in Eq.~\eqref{eq:scalar_field_freqdomain} as
\begin{equation}
    \psi_{lmn}(r) = -4\pi\bigg\{ C^{\rm{in}}_{lmn}(r){} R^{\rm{in}}_{lmn}(r) +  C^{\rm{up}}_{lmn}(r) R^{\rm{up}}_{lmn}(r)\bigg\},
\end{equation}
where
\begin{align}
     C^{\rm{in}}_{lmn}(r) = \int_{r}^\infty \, dr' \dfrac{{} R^{\rm{up}}_{lmn}(r')}{W_{lmn}} r'^2  \rho_{lmn}(r') \, ,\\
     C^{\rm{up}}_{lmn}(r) = \int_{2M}^r \, dr' \dfrac{ R^{\rm{in}}_{lmn}(r')}{W_{lmn}} r'^2 \rho_{lmn}(r') \, .
\end{align}
Taking the limit $r\to\infty$, we notice that $C^{\rm{in}}_{lmn}=0$ and only $C^{\rm{up}}_{lmn}(r)$ needs to be evaluated. Hence, the asymptotic solution takes the simplified form
\begin{equation}
    \psi_{lmn}(r\to\infty) = -4\pi \tilde{R}^{\rm{up}}_{lmn}\int_{2M}^\infty \, dr' \dfrac{ R^{\rm{in}}_{lmn}(r')}{W_{lmn}} r'^2 \rho_{lmn}(r'),
\end{equation}
where $\tilde{R}^{\rm{up}}_{lmn}$ is 
\begin{equation}
    \tilde{R}^{\rm{up}}_{lmn} = C^{{\rm{trans}}}\dfrac{e^{i\omega r_*}}{r},
\end{equation}
and $r_*$ is the usual tortoise coordinate defined as
\begin{equation}
    r_* = r + 2GM\log\left|\dfrac{r}{2GM} - 1\right|\,.
\end{equation}
Further, $C^{{\rm{trans}}}$  is the transmission coefficient. The latter can be computed using the MST formalism, which automatically fulfills the boundary condition we are interested in. A detailed discussion on how to compute these coefficients can be found in \cite{Sasaki:2003xr} and its expression for a scalar charge can be found in Eq.~(5.10) in \cite{Bini:2016egn}. 

By using standard notation given, for instance, in \cite{Sasaki:2003xr,Pound:2021qin}, we re-write the asymptotic field as
\begin{equation}\label{eq:asymptotic_field}
    \psi_{lmn} (r\to\infty) = -4\pi e^{i\omega r_*} \dfrac{Z^{\infty}_{lmn}}{r}\,,
\end{equation}
and its time domain counterpart can be written simply as 
\begin{equation}
    \psi_{lm}(t,r\to\infty) = -4\pi \sum_n e^{-i\omega t} e^{i\omega r_*} \dfrac{Z^{\infty}_{lmn}}{r}\,,
\end{equation}
where the $Z^{\infty}_{lmn}$ are the frequency domain amplitudes defined as
\begin{equation}
    Z^{\infty}_{lmn} = C^{{\rm{trans}}}\int_{2M}^\infty \, dr' \dfrac{ R^{\rm{in}}_{lmn}(r')}{W_{lmn}} r'^2 \rho_{lmn}(r')\,.
\end{equation}

We can now directly define the outgoing energy flux. The simplest procedure that one can use is the same as that used in \cite{Bianchi:2024rod}, but here we are generalizing their discussion to the eccentric case.
Let us start from the definition of the energy-momentum tensor 
\begin{equation}
    T_{\mu\nu} = \dfrac{\mu^2}{8\pi}\left\{\partial_\mu\psi\partial_\nu\psi^*+\partial_\nu\psi\partial_\mu\psi^* - g_{\mu\nu}\partial^\lambda\psi\partial_\lambda\psi^*\right\},
\end{equation}
where $\psi$ is a time domain complex scalar field and the superscript $*$ labels the complex conjugate. The energy per unit volume is then defined as
\begin{equation}
    \dfrac{d^2E^\infty}{d\Omega dt} = \lim_{r\to\infty}r^2 \,T^r{}_t\,,
\end{equation}
and we can write 
\begin{equation}
    \dfrac{d^2E^\infty}{d\Omega dt} = \lim_{r\to\infty}\dfrac{\mu^2 r^2}{8\pi} \, f(r) \left\{ \partial_r\psi\partial_t\psi^* + \partial_t\psi\partial_r\psi^* \right\}\,.
\end{equation}

By expanding the scalar field in the usual basis of spherical harmonics, we find
\begin{align}\nonumber
  \dfrac{dE^\infty}{dt}= \lim_{r\to\infty}\dfrac{\mu^2}{8\pi}\sum_{l,m} r^2 f(r) \bigg\{\partial_r\psi_{lm}\partial_t\psi^*_{lm} + \text{c.c}\bigg\}\,,
\end{align}
where we have already integrated over the sphere and leveraged the orthogonality property for $Y_{l,m}(\theta,\phi)$. In this expression, $\psi_{lm}$ (and its complex conjugate), is indeed the time domain solution of the KG equation we derived in Eq.~\eqref{eq:scal_field_timedom}

The derivatives of the asymptotic fields are simply
\begin{subequations}
    \begin{align}
    \partial_t\psi_{lm}(t,r\to\infty) & = -4\pi \sum_n (-i\omega) e^{-i\omega t} e^{i\omega r_*} \dfrac{Z^{\infty}_{lmn}}{r}\,, \\
    \partial_r\psi_{lm}(t,r\to\infty) & = -4\pi \sum_n  e^{-i\omega t} Z^{\infty}_{lmn} \partial_r \left(\dfrac{e^{i\omega r_*}}{r}\right)\,, 
\end{align}
\end{subequations}
where 
\begin{equation}
    \partial_r\left(\dfrac{e^{i\omega r_*}}{r}\right) = -\dfrac{e^{i\omega r_* }}{r^2} + i\omega\partial_r r_* \dfrac{e^{i\omega r_*}}{r}\,,
\end{equation}
with $\partial_r r_*=1 + \dfrac{2GM}{r}\dfrac{1}{f(r)} $. Because we only care about the $1/r$ behavior, we then have 
\begin{align}\nonumber
    \partial_r\psi_{lm}(t,r\to\infty) & = -4\pi \sum_n i\omega  e^{-i\omega t} Z^{\infty}_{lmn} \dfrac{e^{i\omega r_*}}{r} \\&= - \partial_t \psi_{lm}(t,r\to\infty)\,.
\end{align}

Putting everything together, the flux assumes then the following compact expression
\begin{equation}
    \dfrac{dE^\infty}{dt} = - \dfrac{\mu^2}{4\pi} \sum_{l,m} |\dot\psi_{lm}|^2  ,
\end{equation}
where the dot is the coordinate time derivative of the scalar field.

The procedure described above can be reiterated for the angular momentum flux, by looking at the $(r,\phi)$ component of the stress energy tensor and we find 
\begin{align}
    \dfrac{dL^\infty}{dt} & = -\dfrac{\mu^2}{8\pi} \sum_{l,m} m\left\{ \dot{\psi}_{lm}\psi^* _{lm} + \text{c.c} \right\}  \\&= -\dfrac{\mu^2}{4\pi} \sum_{l,m}m\,\text{Re}\left\{ \dot{\psi}_{lm}\psi^* _{lm} \right\}\,.
\end{align}

In order to compute explicit expressions for the fluxes, it is crucial to perform the sum over $(l,m)$. However, in this case, the sum is finite and we only need to take into account a finite number of modes, namely $l=0,...,4$.
Explicit expressions for the fluxes in units of $\varepsilon$ are given below
\begin{align}\nonumber\label{res:energy_flux}
    \dfrac{dE^\infty}{dt}  &= -\frac{y^4}{3}+\frac{2 y^5}{3}-\frac{2}{3} \pi  y^{11/2}+\frac{10 y^6}{3} \\ \nonumber 
    &+e^2
   \left(-y^4+\frac{7 y^5}{3}-\frac{14}{3} \pi  y^{11/2}+\frac{92 y^6}{3}\right)\\ 
   &+e^4
   \left(-\frac{15 y^4}{8}  +\frac{17 y^5}{3}-\frac{67}{4} \pi 
   y^{11/2}+\frac{945 y^6}{8}\right) \,, 
\end{align}
\begin{align}\nonumber\label{res:angmom_flux}
     \dfrac{dL^\infty}{dt}  &= -\frac{y^{5/2}}{3}+\frac{2 y^{7/2}}{3}-\frac{2 \pi  y^4}{3}+\frac{10
   y^{9/2}}{3} \\ \nonumber 
   & +e^2 \left(-\frac{y^{5/2}}{3}+\frac{7 y^{7/2}}{6}-2 \pi 
   y^4+\frac{59 y^{9/2}}{3}\right) \\  
   & +e^4
   \left(-\frac{y^{5/2}}{3}+\frac{5 y^{7/2}}{3}-\frac{15 \pi  y^4}{4}+\frac{405
   y^{9/2}}{8}\right)\,,
\end{align}
where, to be consistent, we kept the same number of PN-coefficients for both fluxes. It is important to recall that this expansion must always be done as a \textit{last} step of the calculation in order to avoid a polynomial growth in time that would generate divergent fluxes. 

\begin{figure*}[t]
  \centering
  \includegraphics[width=\linewidth]{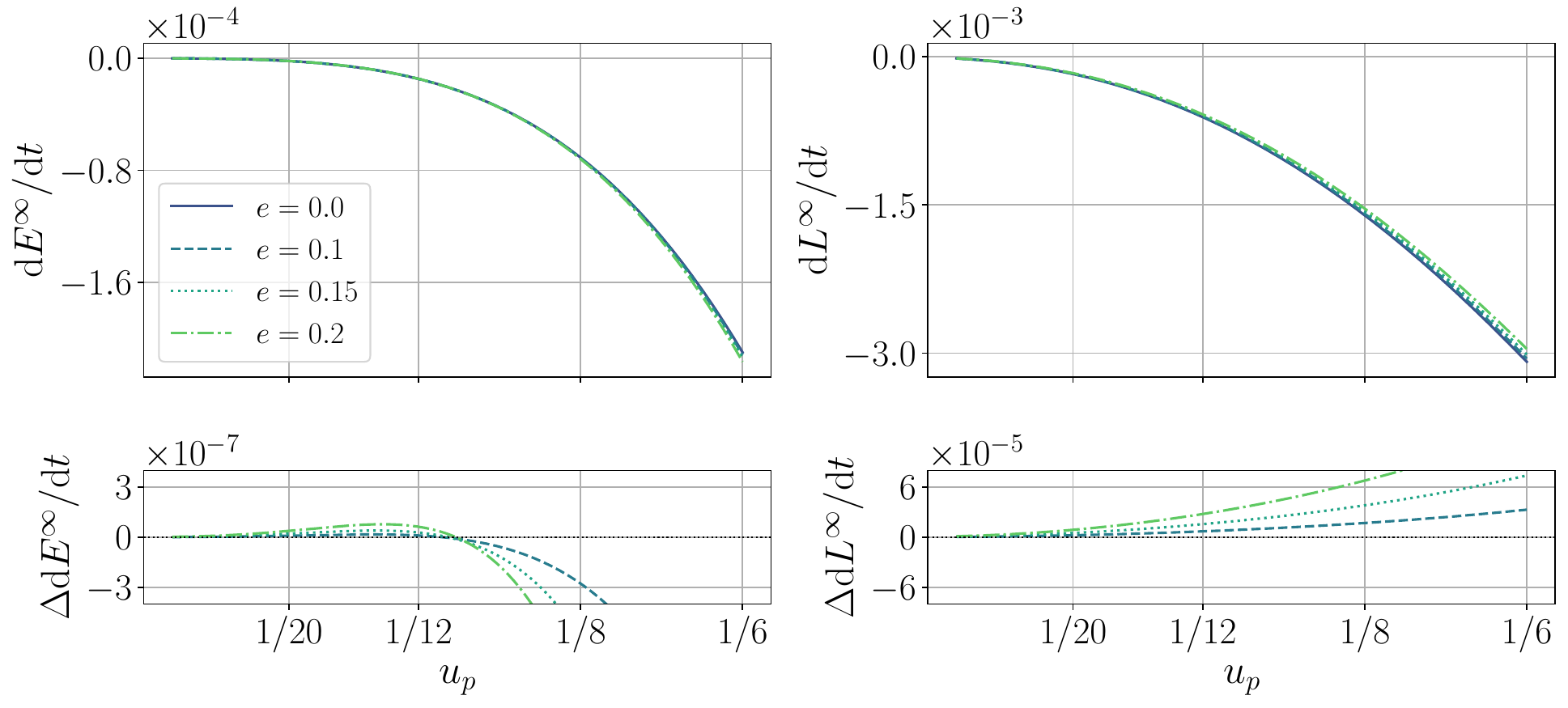}
  \caption{We show here the energy and angular momentum flux in the first row, while in the second we show the difference with respect to the circular case. The fluxes do not differ significantly for the different values of eccentricity we chose.}
  \label{fig:fluxes}
\end{figure*}

The usual comparisons against the circular orbit are presented in Figure \ref{fig:fluxes}. \textcolor{black}{We further provide plots of the quantity}
\begin{equation}
\Delta\left(\dfrac{d\xi^\infty}{dt}\right) = \dfrac{d\xi^\infty}{dt} - \dfrac{d\xi^\infty}{dt}\bigg|_{e\to0}\,,
\end{equation}
where $\xi=\{E,L\}$ are presented in Eqs.~\eqref{res:energy_flux} and~\eqref{res:angmom_flux}, respectively. 
We see that in this case the difference with the circular limit is $\sim 10^{-7}$ for the energy flux in the domain of $e\leq0.2$ and $u_p\geq1/8$.  In the same range of $(u_p,e)$, the angular momentum flux is two orders of magnitude larger.

\section{Comparing with Scalar-Tensor Theories}\label{sec_6}

Let us compare our Eqs.~\eqref{res:energy_flux}-\eqref{res:angmom_flux} with the scalar fluxes presented in Eqs.~(4.12) of Ref.~\cite{Trestini:2024zpi}. The relation between the two results is non-trivial since the orbital parametrization is different: $(y,e)$ in our case and $(x,e_t)$ in~\cite{Trestini:2024zpi}, where this choice has been made to recover trivially the circular limit.

The first step is to change variables from $(y,e)$ to $(y,e_t)$, where $e_t$ is the time eccentricity which is related to the Darwin eccentricity $e$ via Eq.~(4.38) of \cite{Forseth:2015oua} up to third post-Newtonian order. For our purposes, we don't need to include the full expression and the 1PN correction is sufficient in the small eccentricity limit for the comparison. Hence, the relation between the two eccentricities is
\begin{equation}
    e = e_t(1+3y)\,.
\end{equation}

By subsequently changing variables in our fluxes and keeping only the first two terms in the PN expansion, we get
\begin{align}\nonumber \label{eq:en_flux_comp}
   \tilde{\dfrac{dE^\infty}{dt}} & = \dfrac{q^2\mu^2}{M^2} \Bigg\{-\frac{y^4}{3}+\frac{2
   y^5}{3}-\left(y^4+\frac{11
   y^5}{3}\right)
   e_t^2 \\
   &-\left(\frac{15
   y^4}{8} +\frac{101 y^5}{6}\right)
   e_t^4\Bigg\}\,,
\end{align}
\begin{align}\nonumber\label{eq:angmom_flux_comp}
   \tilde{\dfrac{dL^\infty}{dt}} & = \dfrac{q^2\mu^2}{M} \Bigg\{-\frac{y^{5/2}}{3}+\frac{2
   y^{7/2}}{3}-\left(\frac{y^{5/2}}{3}+\frac{5 y^{7/2}}{6}\right)
   e_t^2 \\
   &-\left(\frac{y^{5/2}}{3}+\frac{7
   y^{7/2}}{3}\right) e_t^4\Bigg\}\,,
\end{align}
where we have re-introduced the physical dimensions. We also point out that these expressions are equivalent to the averaged fluxes on a single orbit, i.e.
\begin{equation}
    \left<\tilde{\dfrac{dE^\infty}{dt}} \right> \equiv \dfrac{1}{\Delta T_r} \int_{0}^{\Delta T_r} dt \tilde{\dfrac{dE^\infty}{dt}}\,,  
\end{equation}
\begin{equation}
    \left<\tilde{\dfrac{dL^\infty}{dt}} \right> \equiv \dfrac{1}{\Delta T_r} \int_{0}^{\Delta T_r} dt \tilde{\dfrac{dL^\infty}{dt}}\, , 
\end{equation}
which are the quantities computed in \cite{Trestini:2024zpi}.

As previously stated, the results presented in Ref.~\cite{Trestini:2024zpi}, in particular Eqs.~(4.12), are written in terms of the couple $(x,e_t)$, where $x$ differs from $y$ by terms proportional to the symmetric mass ratio $\nu$, namely $x = y(1+2\nu/3 + O(\nu^2))$ for small $\nu$.

In order to produce a consistent check between the two results we need to perform the following identification on the notation used in Ref.~\cite{Trestini:2024zpi}
\begin{align}\nonumber
    m_1 = M\,,& m_2 = \mu \,, m = m_1 + m_2 = M + \mu\,, \\
   &\nu = \dfrac{\mu}{M} \,, \delta^2 = 1-4\dfrac{\mu}{M}\,, s_1 = \dfrac{1}{2}\,,
\end{align}
where $s_1$ is the sensitivity of the Schwarzschild black hole to variations of the scalar field. Further, the sensitivity and the scalar charge are also related via
\begin{equation}
    q = \dfrac{1-2s_2}{\sqrt{3+2\omega_0}},
\end{equation}
see e.g. \cite{Palenzuela:2013hsa}. Here, $\omega_0$ is a ST parameter and $s_2$ is the sensitivity related to the second body, the scalar particle in our case\footnote{In PN theory there is no concept of primary or secondary object in the binary system. One could indeed reverse the definition of $s_1$ and $s_2$ without any loss of generality.}. It is worthwhile to mention that we fix $s_1$ to be 1/2, i.e. its GR value, because in our investigation the central black hole is not endowed with a scalar charge.

The triple $(s_1,s_2,\omega_0)$ defines a set of unambiguous parameters that we need to use to perform the comparison. In addition to this, we also need to expand Eqs.~(4.12) in Ref.~\cite{Trestini:2024zpi} for small eccentricities and small symmetric mass ratio $\nu$. This last expansion is crucial to make sure that we select only the first order in SF from the ST fluxes.

Once all these manipulations have been done, one can finally rewrite the fluxes of \cite{Trestini:2024zpi} as
\begin{align}\nonumber
    \dfrac{dE^{\rm{ST}}}{dt}&=\frac{\nu^2\phi_0 \left(1-2 s_2^0\right){}^2}{ \left(3+2 \omega _0\right)} \Bigg\{ -\frac{y^4}{3}+\frac{2 y^5}{3}\\
    & +\left(-y^4-\frac{11 y^5}{3}\right) e_t^2+\left(-\frac{15
   y^4}{8}-\frac{101 y^5}{6}\right) e_t^4 \Bigg\}\,,
\end{align}
\begin{align}\nonumber
    \dfrac{dL^{\rm{ST}}}{dt}&=\frac{\nu^2 M \left(1-2 s_2^0\right){}^2}{ \left(3+2 \omega _0\right)} \Bigg\{ -\frac{y^{5/2}}{3}+\frac{2 y^{7/2}}{3}\\
    & +\left(-\frac{y^{5/2}}{3}-\frac{5 y^{7/2}}{6}\right)
   e_t^2+\left(-\frac{y^{5/2}}{3}-\frac{7 y^{7/2}}{3}\right) e_t^4 \Bigg\}\,,
\end{align}
and one can immediately see the equivalence of the above relations with those shown in Eqs.~\eqref{eq:en_flux_comp}\, and~\eqref{eq:angmom_flux_comp}, by comparing the content in the curly brackets.

Besides, by substituting the definition of the scalar source $q$ in terms of sensitivities and $\omega_0$, we see that also the prefactor is in agreement with our findings up to a conventional multiplicative factor $\phi_0$. This comes from the normalization of the scalar field used in ST theory at infinity and can be straightforwardly set to 1. In the context of ST theory, $\phi_0$ it usually comes from the requirement of having a constant value for the field at spatial infinity and it can be interpreted as a boundary condition when solving the scalar field equation.

\section{Discussion and Conclusions}

In this work, we presented analytical results for a scalar charge on an eccentric orbit around a Schwarzschild black hole, using the scalar self-force (SSF) framework. The scalar field evolution is governed by the Klein–Gordon equation, which we solved in Sec.~\ref{sec_3} by combining Post-Newtonian (PN) and Mano–Suzuki–Takasugi (MST) solutions. The resulting expressions preserve the periodic structure required by the orbital motion. This approach enabled us to compute both the SF components acting on the particle in Sec.~\ref{sec_4} and the scalar radiation emitted by the system in Sec.~\ref{sec_5}. 

Triggered by the recent interest in characterizing binary systems in the context of scalar-tensor (ST) theories, we also performed a consistency check of our SSF results by comparing them with the analytical fluxes computed in PN approximation for ST theories (see Eq.~(4.12) of Ref.~\cite{Trestini:2024zpi}) in Sec.~\ref{sec_6}. By finding perfect agreement, we showed how SSF can consistently mimic the results obtained in these works once they are reduced to first-order mass-ratio corrections.

It is important to note that, in order to provide a \textit{complete} identification of these calculations with the ST results, we would need to include the backreaction of the massive scalar field on the background geometry, which is not taken into account in our calculations. However, recent works in this direction have been performed in \cite{Barsanti:2022ana,DellaRocca:2024pnm,Speri:2024qak,Spiers:2023cva} to investigate these classes of models with the SF approach. 

Since our work is completely analytical, it would be interesting to compare our expressions to numerical simulations, in order to check which region of the parameter space can be described sufficiently well within our approach. Our results can be used to inform the scalar sector of EOB potentials for ST theories (see Ref.~\cite{Jain:2022nxs}) from SSF analytical calculations, thereby expanding the possibility to make more precise tests of GR. This would be particularly important because such EOB potentials are used to produce catalogs of simulated waveforms. These catalogs are then employed, once a GW event occurs, and to extract the physical parameters of the astrophysical sources generating the gravitational signal.

\section*{Acknowledgments}
The authors acknowledge  the Istituto Italiano di Fisica Nucleare (INFN) iniziative specifiche QGSKY and MOONLIGHT2.  
SC thanks the Gruppo Nazionale di Fisica Matematica (GNFM)  of Istituto Nazionale di Alta Matematica (INDAM) for the support. 
NM warmly thanks Miguel Zumalac\'arregui for insightful comments on the manuscript.
NM and DU gratefully acknowledge the Max Planck Institute for Gravitational Physics (Albert Einstein Institute) in Potsdam, where this work was originally conceived.
DU thanks Donato Bini, Andrea Geralico, Chris Kavanagh and David Trestini for helpful discussions and clarifications. 
This paper is based upon work from COST Action CA21136, 
Addressing observational tensions in cosmology with systematics and fundamental physics (CosmoVerse) supported by COST (European Cooperation in Science and Technology). 

\bibliography{bibliography}

\end{document}